\documentclass[draftclsnofoot,12pt,onecolumn,oneside,compress]{IEEEtran}
\IEEEoverridecommandlockouts
\ifCLASSINFOpdf
\else
\fi
\usepackage{cite}
\usepackage{amsmath,graphicx,amssymb,color,textcomp,amsfonts,subfigure,enumerate}
\usepackage{algorithm,algorithmic,extarrows}
\usepackage{amsthm}
\usepackage{bm}
\usepackage{balance}
\usepackage{color}

\usepackage[unicode=true,
 bookmarks=false,bookmarksnumbered=true,bookmarksopen=false,bookmarksopenlevel=1,
 breaklinks=false,pdfborder={0 0 0},pdfborderstyle={},backref=false,colorlinks=false]
 {hyperref}
\hypersetup{pdftitle={Title},
 pdfauthor={Admin},
 pdfpagelayout=OneColumn, pdfnewwindow=true, pdfstartview=XYZ, plainpages=false}

\begin{document}
\title{Model-Driven Deep Learning for Non-Coherent Massive Machine-Type Communications}

\author{Zhe~Ma, Wen~Wu,~\IEEEmembership{Senior~Member,~IEEE,} Feifei~Gao,~\IEEEmembership{Fellow,~IEEE,} and Xuemin (Sherman) Shen,~\IEEEmembership{Fellow,~IEEE}
\thanks{Z. Ma and F. Gao are with the Institute for Artificial Intelligence Tsinghua
University, State Key Lab of Intelligent Technologies and Systems, Beijing
National Research Center for Information Science and Technology, Department of Automation, Tsinghua University, Beijing 100084, China (e-mail: maz16@mails.tsinghua.edu.cn; feifeigao@ieee.org).}
\thanks{W. Wu is with the Frontier Research Center, Peng Cheng Laboratory, Shenzhen, Guangdong 518055, China (email: wuw02@pcl.ac.cn).}
\thanks{X. Shen is with the Department of Electrical and Computer Engineering, University of Waterloo, Waterloo, ON N2L 3G1, Canada (e-mail: sshen@uwaterloo.ca).}
}

\maketitle
\vspace{-15mm}
\begin{abstract}
In this paper, we investigate the joint device activity and data detection in massive machine-type communications (mMTC) with a one-phase non-coherent scheme, where data bits are embedded in the pilot sequences and the base station simultaneously detects active devices and their embedded data bits without explicit channel estimation. Due to the correlated sparsity pattern introduced by the non-coherent transmission scheme, the traditional approximate message passing (AMP) algorithm cannot achieve satisfactory performance. Therefore, we propose a deep learning (DL) modified AMP network (DL-mAMPnet) that enhances the detection performance by effectively exploiting the pilot activity correlation. The DL-mAMPnet is constructed by unfolding the AMP algorithm into a feedforward neural network, which combines the principled mathematical model of the AMP algorithm with the powerful learning capability, thereby benefiting from the advantages of both techniques. Trainable parameters are introduced in the DL-mAMPnet to approximate the correlated sparsity pattern and the large-scale fading coefficient. Moreover, a refinement module is designed to further advance the performance by utilizing the spatial feature caused by the correlated sparsity pattern.
Simulation results demonstrate that the proposed DL-mAMPnet can significantly outperform traditional algorithms in terms of the symbol error rate performance.
\end{abstract}

\begin{IEEEkeywords}
Massive machine-type communication (mMTC), non-coherent transmission, grant-free random access, deep learning, model-driven.
\end{IEEEkeywords}

\IEEEpeerreviewmaketitle

\section{Introduction}
To embrace the forthcoming era of Internet of Things (IoT), the 3rd Generation Partnership Project (3GPP) has specified massive machine-type communications (mMTC) as one of the three main service classes for fifth-generation (5G) network and beyond \cite{1}. In a typical mMTC scenario,  a massive number of IoT devices are required to establish uplink-dominated communication with a single base station (BS) \cite{2}. The uplink transmission is usually sporadic and has a short packet size, so only a small and random subset of devices are active for a short while~\cite{3}-\cite{4}.  As a result, conventional grant-based random access protocols are inappropriate for the mMTC scenarios. To better support mMTC services, one potential solution is to develop novel multiple-access schemes that can accomplish user activity and data detection in a timely and accurate manner.

Grant-free (GF) random access is a promising solution for mMTC and IoT, as it eliminates the signaling overhead required for the coordination between the BS and massive devices \cite{5}. In the GF-random access, the user activity and data detection are usually conducted through a two-phase coherent scheme. Specifically, each activated device directly transmits a unique pilot sequence followed by data packets without a prior scheduling assignment. After receiving the superimposed signal from these devices, the BS first detects the active devices and estimates the channel, based on which the corresponding transmitted data bits are then decoded. However, due to the massive number of devices, it is impossible to assign orthogonal pilot sequences to each device, which inevitably leads to collisions among devices and results in performance degradation~\cite{6}. Thanks to the sporadic mMTC traffic pattern, the device activity detection and channel estimation can be formulated as a compressed sensing (CS) problem \cite{7}. Consequently, various CS techniques have been considered for device detection in mMTC, and they have been shown to outperform traditional methods by mitigating pilot contamination~\cite{8}-\cite{10}. Nevertheless, the two-phase coherent scheme incurs non-negligible overhead for channel training. Thus it may not be suitable for mMTC where devices usually transmit small packets intermittently, prompting researchers to consider the non-coherent schemes \cite{101}-\cite{14}.

Several existing works have attempted to investigate the one-phase non-coherent scheme~\cite{11}-\cite{14}. In contrast to the coherent scheme, explicit channel estimation is not required in the non-coherent scheme. The intuition behind the one-phase non-coherent scheme is to allocate multiple distinct pilot sequences to each device. When transmitting, each device selects only one pilot sequence based on its data, and the BS detects the user activity and data jointly by determining which pilot sequence is received. The paper \cite{11} proposes a novel method for embedding 1 bit in pilot sequences, which outperforms the two-phase coherent scheme. The work \cite{12} considers the case when multiple bits are embedded and conducts joint user activity and data detection using the approximate message passing (AMP) algorithm. In \cite{13}, a modified-AMP algorithm is proposed, where the soft-thresholding function is utilized to decide on one of the possible pilot sequences while suppressing the other ones. In~\cite{14}, a covariance-based detection scheme is developed to acquire the indices of the transmitted pilot sequences. However, all the aforementioned works assume that the activity of each pilot sequence is independently and identically distributed. Although the i.d.d. assumption produces an analytically tractable solution, it neglects the correlation among the pilot sequence activity in each user and thus may not be optimal. In this work, we investigate the possibility of applying the deep learning method to explore the correlation structure of the sparsity pattern and improve the joint user activity and data detection.

Thanks to the strong capability of solving intricate and intractable problems, machine learning has become a favorable research topic for future wireless communications \cite{141}-\cite{21}. In particular, as a major branch in machine learning, deep learning has been extensively investigated for signal detection \cite{17}, channel estimation \cite{18}, and constellation design \cite{19} to improve performance while reducing computational complexity. Among vast techniques that employ deep learning in wireless communication, the ``deep unfolding" method that unfolds iterative algorithms into deep neural networks (DNN) is especially attractive\cite{20}.
By incorporating communication expert knowledge into DNN, ``deep unfolding" inherits the mathematical models of classic algorithms and enables the interpretation of network topology design \cite{21}. Meanwhile, by exploiting the powerful learning capability of DL, ``deep unfolding" compensates the imperfections resulting from the inaccuracy of the model and predetermined parameters.

Motivated by existing works, we propose a model-driven DL algorithm, namely DL-modified AMP network (DL-mAMPnet), for the joint device activity and data detection in mMTC with single-phase non-coherent scheme. DL-mAMPnet is constructed by unfolding the AMP algorithm while adding trainable parameters and a refinement module to explore the correlated sparsity pattern of the pilot sequence activity. Simulation results validate the superior symbol error rate (SER) performance of the proposed DL-mAMPnet. The main contributions can be summarized as follows.

\begin{itemize}
\item We formulate the joint device activity and data detection in mMTC with single-phase non-coherent scheme as a hierarchical CS problem with two-level sparsity, where the device activity sparsity and transmitted pilot sequence sparsity are modeled as the system-level sparsity and the device-level sparsity, respectively.
\item We propose an AMP-based algorithm to solve the formulated CS problem. On this basis, we discuss the limitations of the AMP-based algorithm, which serves as the underlying motivation for designing the DL-based algorithm.
\item We propose a DL-based algorithm, termed DL-mAMPnet, to conduct the device activity and data detection jointly. DL-mAMPnet is composed of multiple AMP layers and one refinement module. The AMP layers are obtained by unfolding the AMP algorithm into a feedforward DNN, where trainable parameters are introduced to compensate for the inaccurate i.d.d model of the traditional AMP algorithm. The refinement module exploits the unique spatial feature of the two-level sparsity structure to refine the output of the AMP layers.
\end{itemize}

The remainder of the paper is organized as follows. In Section II, we present the system model and briefly introduce the non-coherent scheme. In Section III, we formulate a hierarchical CS problem with two-level sparsity and correspondingly derive an AMP-based algorithm. In Section IV, we elaborate the structure of the proposed DL-mAMPnet. In Section V, we present the parameter initialization and training method of the proposed DL-mAMPnet. Simulation results are presented in Section VI, and conclusions are made in Section VII.

\emph{Notations:} We use normal lower-case, bold lower-case, and bold upper-case letters to denote scalars, vectors, and matrices, respectively. For matrix $\bm{X}$, $\bm{X}^{T}$ denotes its transpose, $\bm{X}^{H}$ denotes its Hermitian transpose, $|\bm{X}|$ denotes its determinant, and $||\bm{X}||_F$ denotes its Frobenius norm. For vector $\bm{x}$, $||\bm{x}||_p$ denotes its ${l}_p$-norm. $\mathbb{E}\{ \cdot \}$ denotes the expectation operation. $\mathbb{R}^{M \times N}$ and $\mathbb{C}^{M \times N}$ denote the $M \times N$ dimensional real space and complex space, respectively. $\mathcal{CN}(\boldsymbol{\mu},\boldsymbol{\Sigma})$ denotes the multivariate complex Gaussian distribution with mean $\boldsymbol{\mu}$ and covariance $\boldsymbol{\Sigma}$.

\section{System Model}
\subsection{Uplink Massive Access Scenario in mMTC Systems}
We consider a typical uplink massive access scenario in mMTC systems, where a set of randomly distributed single-antenna devices, denoted by $\mathcal{N}=\{1,\cdots,N\}$, communicate with a BS equipped with $M$ antennas. The uplink channel from device $n$ to the BS is denoted by $\bm{h}_n \in \mathbb{C}^{M \times 1}$ and modeled as
\begin{equation}\label{sysm1}
\bm{h}_n=\sqrt{\beta_n}\bm{g}_n, \forall n \in \mathcal{N},
\end{equation}
where $\beta_n$ is the large-scale fading component and $\bm{g}_n$ denotes the small-scale fading component. We assume $\bm{g}_n$ is distributed as $\mathcal{CN}(\bm{0},\bm{I}_M)$, and accordingly we have $\bm{h}_n \sim \mathcal{CN}(\bm{0},\beta_n\bm{I}_M)$. This paper adopts a block-fading channel model, where $\bm{h}_n$ remains unchanged within channel coherence time but is independent from block to block.

Due to the sporadic activity pattern of mMTC, only a small fraction of devices are active in each block. We assume that the devices are synchronized, and each device independently decides whether to access the channel with probability $\epsilon$ in each block. Consequently, the device activity indicator for device $ n \in \mathcal{N}$ is defined as
\begin{equation}\label{sysm2}
\alpha_n=
\begin{cases}
1,& \text{if device} \; n \; \text{is active},\\
0,& \text{otherwise},
\end{cases} \;\;
\end{equation}
where $\text{Pr}(\alpha_n=1)=\epsilon$ and $\text{Pr}(\alpha_n=0)=1-\epsilon$. We further define the set of active devices within a block as
\begin{equation}\label{sysm3}
\mathcal{K}=\{n \in \mathcal{N} : \alpha_n =1 \},
\end{equation}
and the number of active devices is $K=|\mathcal{K}|$.  The received signal $\bm{y} \in \mathbb{C} ^{M \times 1}$ at the BS is given by
\begin{equation}\label{sysm4}
\bm{y}=\sum_{n \in \mathcal{N}} \alpha_n \bm{h}_n x_n + \bm{n} = \sum_{k \in \mathcal{K}}  \bm{h}_k x_k + \bm{n},
\end{equation}
where $x_n \in \mathbb{C}$ is the transmitted signal of device $n$, and $\bm{n} \in \mathbb{C}^{M \times 1}$ is the additive white Gaussian noise (AWGN) distributed as $\mathcal{CN}(\bm{0}, \sigma^2 \bm{I}_M)$.

\subsection{One-Phase Non-Coherent Scheme}
To successfully transmit the messages of the active devices, two schemes have been proposed in the literature, namely the two-phase coherent scheme and the one-phase non-coherent scheme. The two-phase coherent scheme divides each coherence block into two contiguous phases.  In the first phase, the active devices send their pilot sequences to the BS synchronously, and the BS jointly detects the device activity, i.e., $\alpha_n$, as well as their corresponding channels, i.e., $\bm{h}_n$, $\forall n \in \mathcal{K}$. In the second phase, the active devices send their messages to the BS using the remaining coherence block, and the BS decodes these messages based on the knowledge of device activity and channels obtained in the first phase.

Unlike the two-phase coherent scheme, the one-phase non-coherent scheme considered in this paper can jointly detect the active devices and the corresponding messages without explicit channel estimation. Specifically, in the non-coherent scheme, the transmitted messages are embedded in the index of the transmitted pilot sequence of each active device.  To this end, each device maintains a unique set of pre-assigned $Q=2^J$ pilot sequences. When a device is active, it sends a $J$-bit message by transmitting one sequence from the set. By detecting which sequences are received, the BS acquires both the identity of the active devices as well as the $J$-bit message from each of the active devices. We define the pilot sequences allocated for device $n$ as:
\begin{equation}\label{sysm5}
\bm{S}_n= \{ \bm{s}_{n}^1, \bm{s}_{n}^2,\cdots, \bm{s}_{n}^{Q} \},
\end{equation}
where $\bm{s}_{n}^q= [s_{n_1}^q,s_{n_2}^q,\cdots,s_{n_L}^q]^T \in \mathbb{C}^{L \times 1}, 1 \le q \le Q $, and $L$ is the sequence length. Note that the total number of pilot sequences is usually much larger than the length of pilot sequence (or the length of a coherence block), i.e., $NQ \gg L$. As such, it is impossible to assign mutually orthogonal sequences to all devices. Following the pioneering work \cite{22}, we adopt the random Gaussian sequences in this paper. Specifically, each entry of the pilot sequences is generated from i.i.d complex Gaussian distribution with zero mean and variance $1/L$, i.e., $s_{n_l}^q \sim \mathcal{CN}(0, 1/L)$,  so that each pilot sequence has a unit norm, i.e., $||\bm{s}_{n}^q||_2 =1,$ $\forall n \in \mathcal{N} $ and $q=1, \cdots, Q$.

\begin{figure}[!ht]
  \centering
  \includegraphics[width=10cm]{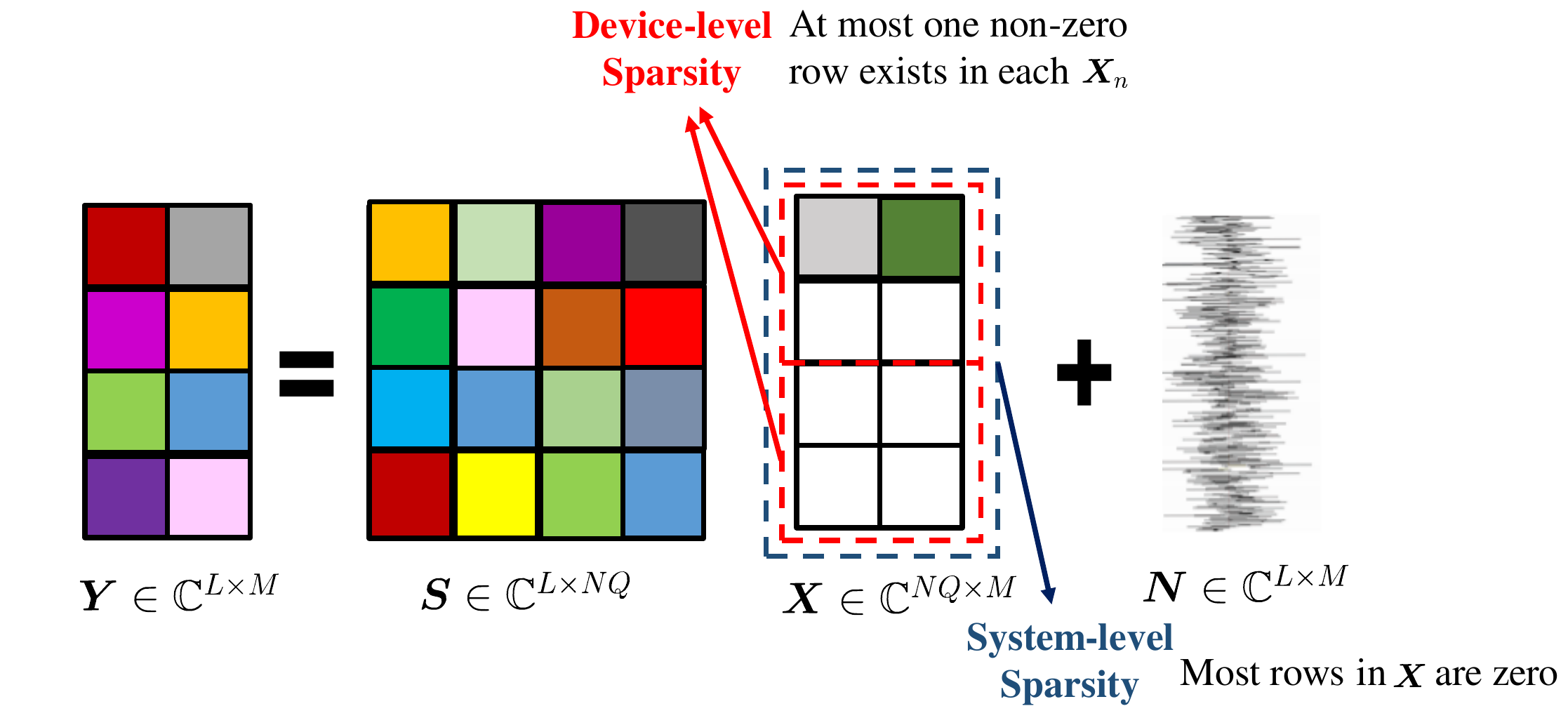}\\
  \caption{Pictorial form of the signal model.}\label{fig1}
\end{figure}

For transmission, each active device selects exactly only one sequence from $\bm{S}_n$ based on its message. Then, the composite received  signal $\bm{Y} \in \mathbb{C}^{L \times M}$ of the non-coherent scheme can  be expressed as
\begin{equation}\label{sysm6}
\bm{Y} =\sum_{n=1}^{N} \sum_{q=1}^{Q} \alpha_{n}^q \bm{s}_{n}^{q} \bm{h}_n^{T} + \bm{N} =\sum_{n=1}^N \bm{S}_n \bm{X}_{n} +\bm{N},
\end{equation}
where $\bm{X}_n=[\alpha_{n}^{1}\bm{h}_n, \alpha_{n}^{2}\bm{h}_n, \cdots, \alpha_{n}^{Q}\bm{h}_n]^{T} \in \mathbb{C}^{Q \times M}$ and $\alpha_{n}^q \in \{0,1\}$ indicates whether or not sequence $q$ of device $n$ is transmitted, with a slight abuse of notation. Recall that each device is active with probability $\epsilon$, we have
\begin{equation}\label{sysm8}
\sum_{q=1}^{Q}\alpha_{n}^{q}=
\begin{cases}
1,& \text{with probability} \; \epsilon;\\
0,& \text{with probability} \; 1-\epsilon.
\end{cases} \;\;
\end{equation}
By further concatenating all sequences of $N$ devices as $\bm{S}=[\bm{S}_1, \bm{S}_2, \cdots, \bm{S}_N ] \in \mathbb{C}^{L \times NQ}$, the received signal in (\ref{sysm6}) can be simplified as
\begin{equation}\label{sysm7}
\bm{Y}=\bm{S} \bm{X} +\bm{N},
\end{equation}
where $\bm{X}=[\bm{X}_1^{T},\bm{X}_2^{T},\cdots,\bm{X}_N^{T}]^{T} \in \mathbb{C}^{NQ \times M}$. The pictorial form of (\ref{sysm7}) is sketched in Fig.~\ref{fig1}, which intuitively shows that $\bm{X}$ has a hierarchical sparse structure. The hierarchical sparse structure comprises two levels of sparsity, including the \emph{system-level sparsity} and the \emph{device-level sparsity}. The system-level sparsity means that most rows in $\bm{X}$ are zero, which is due to the sporadic traffic pattern. The device-level sparsity enforces that there is at most one non-zero row exists in $\bm{X}_n, \forall n$, because each active device only transmits one pilot sequence from its pilot set.

\section{Problem Formulation and AMP-Based Joint Detection Algorithm}
\subsection{Problem Formulation}

Our goal is to detect the binary variable $\alpha_{n}^q$ that indicates both the activity of device $n$ and its transmitted message, which can be achieved by recovering $\bm{X}$ from the received signal $\bm{Y}$. Once $\bm{X}$ is recovered, $\alpha_{n}^q$ can be determined by the rows of $\bm{X}$. Due to the hierarchical sparse structure of $\bm{X}$, such problem is a classic CS problem with known measurement matrix $\bm{S}$. Therefore, we can formulate the problem as follows:

\begin{align}\label{Pfor1}
    \mathcal{P}1: \min_{\bm{X}} \;\;\; &||\bm{Y}-\bm{S}\bm{X}||_{F}^{2} \\
    s.t. \;\;\;  &\sum_{n=1}^{N}\sum_{q=1}^{Q} \mathbb{I}({\bm{X}_n}_{q,:}) \le K,  \label{c1}\\
    &   \sum_{q=1}^{Q} \mathbb{I}({\bm{X}_n}_{q,:}) \le 1, \forall n, \label{c2}
\end{align}
where ${\bm{X}_n}_{q,:}$ is the $q$th row of $\bm{X}_n$ and $\mathbb{I}(\cdot)$ is the indicator function defined as
\begin{equation}\label{Pfor2}
\mathbb{I}(\bm{x})=
\begin{cases}
1,&  \text{if} \;\bm{x}\; \text{has non-zero elements};\\
0,& \text{otherwise}.
\end{cases}
\end{equation}
The constraint (\ref{c1}) comes from the system-level sparsity and the constraint (\ref{c2}) ensures the device-level sparsity.  However, it is challenging to solve $\mathcal{P}1$ directly due to the non-smooth constraints.  Hence, we relax (\ref{c2}) into a $l_{2,1}$-norm regularized least-square problem by replacing the indicator function with $l_2$ norm as \cite{23}
\begin{equation}\label{Pfor3}
\min_{\bm{X}} \;\;\; \frac{1}{2}||\bm{Y}-\bm{S}\bm{X}||_{F}^{2}+ \lambda \sum_{n=1}^{N}\sum_{q=1}^{Q} ||{\bm{X}_n}_{q,:}||_2,
\end{equation}
where $\lambda$ is the tunable parameter that balances the the sparsity of the solution and the mean square error (MSE) $||\bm{Y}-\bm{S}\bm{X}||_{F}^{2}$. Although conventional CS algorithms such as orthogonal matching
pursuit (OMP) and sparse Bayesian learning (SBL) can be directly used to solve (\ref{Pfor3}), they suffer high computational complexity due to the matrix inverse operation, especially in mMTC system with massive devices. In view of this, this paper utilizes the computationally efficient AMP algorithm as the main technique \cite{24}.

\subsection{Review of the AMP Algorithm}
AMP refers to a class of efficient algorithms for statistical estimation in high-dimensional problems such as linear regression and low-rank matrix estimation. The goal of the AMP algorithm is to obtain an estimate of $\bm{X}$ with the minimum MSE based on $\bm{Y}$. Starting with $\bm{X}_0=\bm{0}$ and $\bm{R}_0=\bm{Y}$, the AMP algorithm can be described as follows:
\begin{equation}\label{amp1}
{\bm{X}}_{t+1,n}=\eta_{t,n}(\bm{S}_n^{H}\bm{R}_t+\bm{X}_{t,n}), \forall n,
\end{equation}
\begin{equation}\label{amp2}
\bm{R}_{t+1}=\bm{Y}-\bm{S}\bm{X}_{t+1}+b_t\bm{R}_{t},
\end{equation}
where $t=0,1, \cdots$ is the index of the iteration, $\eta_{t,n}(\cdot)$ is the shrinkage function for device $n$ that shrinks some items of its input to zero, and $\bm{R}_t$ is the corresponding residual. The residual in (\ref{amp2}) is updated with the ``Onsager correction'' term $b_t\bm{R}_{t}$, which substantially improves the performance of the AMP algorithm \cite{25}. Note that $\eta_{t,n}(\cdot)$ is assumed to be Lipschitz-continuous and $b_t$ can be written as
\begin{equation}\label{amp3}
b_t=\frac{1}{L}\sum_{n=1}^{N} \eta_{t,n}^{'}(\bm{S}_n^{H}\bm{R}_t+\bm{X}_{t,n}),
\end{equation}
where $\eta_{t,n}^{'}(\cdot)$ is the first-order derivative of $\eta_{t,n}(\cdot)$. In addition to improving the performance, the Onsager correction also enables the AMP algorithm to be analyzed by a set of state evolution equations in the asymptotic regime \cite{26}. The asymptotic regime is when $L, N \to \infty$,  while their ratio converges to a positive constant, i.e., $N/L \to \rho $ where $\rho \in (0, \infty)$, and while keeping the data length $J$ fixed.  To facilitate the theoretical analysis, this paper considers a certain asymptotic regime where $N \to \infty$, and the empirical distribution of the large-scale fading components $\beta_n$'s converges to a fixed distribution $p_{\beta}$.

Define $\beta \sim p_{\beta}$ and $\bm{X}_{\beta} \in \mathbb{C}^{Q \times M}$ as a random matrix distributed as $(1-\frac{\epsilon}{Q})\prod_{i=1}^{Q}\delta_{\bm{x}_{\beta,i}} + \frac{\epsilon}{Q} \sum_{i=1}^{Q} P_{\bm{h}_{\beta}} \prod_{j \neq i}\delta_{\bm{x}_{\beta,j}}$, where $\delta_{\bm{x}_{\beta,i}}$ is the Dirac delta at zero corresponding to the element $\bm{x}_{\beta,i}$ and $P_{\bm{h}_\beta}$ denotes the distribution $\bm{h}_\beta \sim \mathcal{CN}(\bm{0},\beta\bm{I}_M)$. The state evolution equations can be written as the following recursions for $t\ge0$ \cite{26}
\begin{equation}\label{amp4}
\bm{\Sigma}_{0}=\sigma^2\bm{I}_M+\rho\mathbb{E}_{\beta}\{\bm{X}_{\beta}^H \bm{X}_{\beta} \},
\end{equation}
\begin{equation}\label{amp5}
\bm{\Sigma}_{t+1} =\sigma^2\bm{I}_M+ \rho\mathbb{E}_{\beta}\{(\eta_t(\bm{X}_{\beta}+\bm{V}\bm{\Sigma}_t^{\frac{1}{2}})-\bm{X}_{\beta})^H  (\eta_t(\bm{X}_{\beta}+\bm{V}\bm{\Sigma}_t^{\frac{1}{2}})-\bm{X}_{\beta})\},
\end{equation}
where $\bm{V} \in \mathbb{C}^{Q \times M}$ is a random matrix independent with $\bm{X}_{\beta}$, of which the rows are i.i.d. and each follows the distribution $\mathcal{CN}(\bm{0},\bm{I}_M)$. It can be observed from (\ref{amp1}) and (\ref{amp5}) that applying $\eta_{t,n}(\cdot)$ to $\bm{S}_n^{H}\bm{R}_t+\bm{X}_{t,n}$ is statistically equivalent to applying $\eta_{t,n}(\cdot)$ to $\bm{X}_{t,n}+\bm{V}\bm{\Sigma}_t^{\frac{1}{2}}$. Therefore, the input to the shrinkage function  $\eta_{t,n}(\cdot)$ can be modeled as an AWGN-corrupted signal, i.e.,
\begin{equation}\label{amp6}
\bm{Z}_{t,n}=\bm{X}_{t,n}+\bm{S}_n^{H}\bm{R}_t= \bm{X}_{t,n}+\bm{V}\bm{\Sigma}_t^{\frac{1}{2}},
\end{equation}
In this case, the update given by (\ref{amp1}) is statistically equivalent to a denosing problem, and thus $\eta_{t}(\cdot)$ can also be called ``denoiser''. Hereafter, we use ``shrinkage function" and ``denoiser" interchangeably for convenience.

\subsection{AMP-Based Joint Device Activity and Date Detection Algorithm}
The core idea behind the joint detection algorithm is to first estimate $\bm{X}$ from $\bm{Y}$, based on which $\alpha_{n}^q$ is determined according to the norm of each rows in $\bm{X}$. To this end, we first derive the denoiser $\eta_{t,n}(\cdot)$ under the MMSE-optimal criterion. After that, we observe that $\eta_{t,n}(\cdot)$ exhibits an asymptotic property, which motivates us to design a threshold-based strategy to extract $\alpha_{n}^q$ from $\bm{X}$.

\subsubsection{Derivation of $\eta_{t,n}(\cdot)$}
For notational simplicity, we omit the iteration index $t$ in the following. According to (\ref{amp6}), the likelihood of $\bm{Z}_{n}$ given $\bm{X}_{n}$ takes the form of
\begin{equation}\label{amp7}
P_{\bm{Z}_{n}|\bm{X}_{n}} =\prod_{q=1}^{Q}\frac{\exp (-(\bm{z}_{n}^{q}-\bm{x}_{n}^{q})^{H}\bm{\Sigma^{-1}}(\bm{z}_{n}^{q}-\bm{x}_{n}^{q}))}{\pi^M|\bm{\Sigma}|}.
\end{equation}
Accordingly, the MMSE-optimal denoiser is given by the conditional expectation $\mathbb{E}\{\bm{X}_{n}| \bm{Z}_{n} \}$ and can be expressed as
\begin{equation}\label{amp8}
\eta_{n}(\bm{Z}_{n})=\mathbb{E}\{\bm{X}_{n}| \bm{Z}_{n}\} =[\phi_{n}^{1}\bm{\Omega}_{n}\bm{z}_{n}^{1}, \cdots,\phi_{n}^{Q}\bm{\Omega}_{n}\bm{z}_{n}^{Q}],
\end{equation}
where
\begin{equation}\label{amp9}
\bm{\Omega}_{n}=\beta_n(\beta_n\bm{I}_M+\bm{\Sigma})^{-1},
\end{equation}
\begin{equation}\label{amp10}
\phi_{n}^{q}=\frac{1}{1+\frac{Q-\epsilon}{\epsilon}\exp(M(\psi_{n}-\pi_{n}^{q}))},
\end{equation}
\begin{equation}\label{amp11}
\psi_{n}=\frac{\log(|\bm{I}_M+\beta_n\bm{\Sigma}^{-1}|)}{M},
\end{equation}
and
\begin{equation}\label{amp12}
\pi_{n}^{q}=\frac{{\bm{z}_{n}^{q}}^{H}(\bm{\Sigma}^{-1}-(\bm{\Sigma}+\beta_n\bm{I}_M)^{-1})\bm{z}_{n}^{q}}{M}.
\end{equation}
\emph{Proof:} Please refer to Appendix A.

It is important to realize that the MMSE-optimal denoiser $\eta_{n}(\cdot)$ is rather complicated as it involves the computation of the state evolution matrix $\bm{\Sigma}$, where the matrix multiplication and expectation are needed.  Hence, we simplify $\eta_{n}(\cdot)$ by using the following theorem.

\emph{Theorem 1:} Considering the asymptotic regime where both the number of devices $N$ and the length of the pilot sequences $L$ go to the infinity with their ratio converging to some fixed positive values, i.e., $N/L \to \rho$ where $\rho \in (0, \infty)$, the state evolution matrix $\bm{\Sigma}_t$ always remains as a diagonal matrix with identical diagonal entries after each iteration, i.e.,
\begin{equation}\label{amp13}
\bm{\Sigma}_t=\tau_t^2\bm{I}_M, \; \forall t \ge 0.
\end{equation}

Correspondingly, the signal model given in (\ref{amp6}) reduces to
\begin{equation}\label{amp14}
\bm{Z}_{t,n}=\bm{X}_{t,n}+\bm{S}_n^{H}\bm{R}_t= \bm{X}_{t,n}+\tau_t\bm{V},
\end{equation}
and the MMSE-optimal dnoiser given in (\ref{amp8})-(\ref{amp12}) is simplified as
\begin{equation}\label{amp15}
\eta_{n}(\bm{Z}_{n})=\mathbb{E}\{\bm{X}_{n}| \bm{Z}_{n}\} =[\phi_{n}^{1}\omega_{n}\bm{z}_{n}^{1}, \cdots,\phi_{n}^{Q}\omega_{n}\bm{z}_{n}^{Q}],
\end{equation}
where
\begin{equation}\label{amp16}
\omega_{n}=\frac{\beta_n}{\beta_n + \tau^2},
\end{equation}
\begin{equation}\label{amp17}
\phi_{n}^{q}=\frac{1}{1+\frac{Q-\epsilon}{\epsilon}\exp(M(\psi_{n}-\pi_{n}^{q}))},
\end{equation}
\begin{equation}\label{amp18}
\psi_{n}=\log(1+\frac{\beta_n}{\tau^2}),
\end{equation}
and
\begin{equation}\label{amp19}
\pi_{n}^{q}=\frac{\beta_n{\bm{z}_{n}^{q}}^{H}\bm{z}_{n}^{q}}{\tau^2(\beta_n + \tau^2)M}.
\end{equation}

Finally, $\tau_t^2$ can be obtained using the following recursions for $ t \ge 0$:
\begin{equation}\label{amp20}
\tau_0^2= \sigma^2 + \rho \epsilon \mathbb{E}_{\beta} \{\beta\},
\end{equation}
\begin{equation}\label{amp21}
\tau_{t+1}^2 = \sigma^2 + \rho \sum_{q=1}^{Q}\mathbb{E}_{\beta} \{ \frac{ \phi_{\beta}^{q} \beta \tau_t^2}{\beta+\tau_t^2} \} + \rho \sum_{q=1}^{Q}\mathbb{E}_{\beta} \{\phi_{\beta}^{q}(1-\phi_{\beta}^{q}) \frac{ \beta^2 {\bm{z}_{n}^{q}}^{H}\bm{z}_{n}^{q}}{(\beta+\tau_t^2)^2M} \}.
\end{equation}
We omit the detailed proof here for brevity. Interested readers can refer to theorem 1 in \cite{8}, where a similar derivation is provided. It should be mentioned that the proposed Theorem 1 in this paper is essentially a generalization of Theorem 1 in \cite{8}. When each device is assigned with only one  pilot sequence, i.e., $Q=1$, the proposed Theorem 1 reduces to Theorem 1 in \cite{8}.

\subsubsection{Threshold-Based Strategy}
It can be seen from (\ref{amp15})-(\ref{amp17}) that for large $M$, we have ${\phi}_{n}^{q} \to 1$ if $\pi_{n}^{q}>\psi_{n}$  and  $\phi_{n}^{q} \to 0$ if $\pi_{n}^{q}<\psi_{n}$. The asymptotic behavior of ${\phi}_{n}^{q}$ indicates that it is reasonable to adopt a threshold-based strategy for solution refinement. Meanwhile, considering the device sparsity in (\ref{sysm8}), an element selection operation is necessitated to enforce all the elements except the one with the largest magnitude in each $\bm{X}_n$ to be zeros.  Consequently, the proposed threshold-based strategy should be able to perform the following two operations.

\textbf{Element Selection Operation:} To surely guarantee the sparsity constraint in (\ref{c2}), we choose the largest row in each $\bm{X}_n=[\bm{x}_n^{1}, \bm{x}_n^{2}, \cdots, \bm{x}_n^{Q}]$ and define the index of the largest element as
\begin{equation}\label{rm5}
i^{*}_n=\text{arg}\max_{i} {\bm{x}_n^{i}}^{H} \bm{x}_n^{i}, \forall n \in \mathcal{N}.
\end{equation}

\textbf{Threshold-based Decisive Operation:} After obtaining $i^{*}_n$, the binary variable vector $\bm{\alpha}_n=\{ \alpha_n^1, \cdots,\alpha_n^Q \}$ can be given as
\begin{equation}
\label{rm6}
\bm{\alpha}_n=\left\{
\begin{aligned}
\bm{e}_{i^{*}_n} & , & \text{if} \; \kappa_{n}^{i^{*}_n}> 0; \\
\bm{0} & , &  \; \text{otherwise},
\end{aligned}
\right.
\end{equation}
where $\bm{e}_{i^{*}_n}$ is a one-hot vector of length $Q$ with only the $i^{*}_n$th element equal 1 and the others equal 0, and the corresponding threshold is computed using (\ref{amp18}) and (\ref{amp19}) as
\begin{equation}\label{rm7}
\kappa_{n}^{i^{*}_n}=\frac{{\bm{z}_{n}^{i^{*}_n}}^{H} \bm{z}_{n}^{i^{*}_n}\beta_n }{\tau_t^2(\beta_n+\tau_t^2)M}-\log\left(1+\frac{\beta_n}{\tau_t^2}\right).
\end{equation}

\subsubsection{Limitation} 
Although the traditional AMP-based algorithm can successfully recover $a_{n}^q$ from $\bm{Y}$, it has some inherent limitations: (i) The traditional AMP algorithm implicitly assumes $\bm{X}_n$ has a prior distribution with i.i.d. entries, which neglects the dependencies among the rows of $\bm{X_n}$ imposed by the device-level sparsity; (ii) The calculation of the denoiser $\eta_{t,n}(\cdot)$ and the threshold $\kappa_{n}$ requires the exact value of $\beta_n$, which is costly to obtain in a large-scale mMTC system with massive devices.

\begin{figure*}[!ht]
  \centering
  \includegraphics[width=15cm]{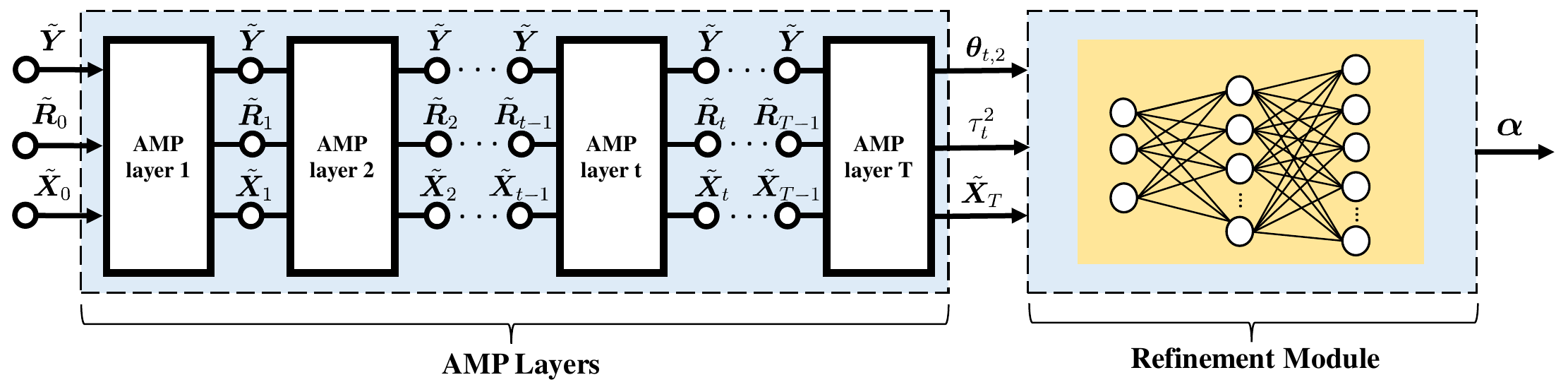}\\
  \caption{Network architecture of the proposed DL-mAMPnet.}\label{fig2}
\end{figure*}

\section{Deep Learning Modified AMP Network}
To address the aforementioned limitations, we propose a deep learning modified AMP network (DL-mAMPnet). The DL-mAMPnet is constructed by unfolding the AMP algorithm into a feedforward DNN, which inherits the mathematical model and structure
of the AMP algorithm, thereby avoiding the requirements for accurate modeling. On this basis, we introduce a few trainable parameters into the DL-mAMPnet to learn the active probability and the large-scale fading. By making the active probability trainable, we compensate for the inaccuracy caused by the i.i.d. assumption in the traditional AMP algorithm. By making the large-scale fading coefficient trainable, we bypass the statistical measurements for the large-scale fadings of massive devices. According to the threshold-based strategy in Section III-C, we further design a refinement module to guarantee the device-level sparsity and obtain the desired $a_{n}^q$.

As depicted in Fig.~\ref{fig2}, the proposed DL-mAMPnet consists of $T$ uniform AMP layers and one refinement module. For the sake of clarity, each part of the DL-mAMPnet is elaborated respectively in the following subsection.

\subsection{Input and Output}
To facilitate the learning process of DL-mAMPnet, the complex matrices need to be converted into the real domain and then vectorized. To do this, we first express (\ref{sysm7}) as
\begin{equation}\label{net1}
\left[
\begin{array}{c}
 \Re(\bm{Y})\\
 \Im(\bm{Y})
\end{array}
\right ]
=
\left[
\begin{array}{cc}
 \Re(\bm{S})&-\Im(\bm{S})\\
 \Im(\bm{S})&\Re(\bm{S})
\end{array}
\right ]
\left[
\begin{array}{c}
\Re(\bm{X})\\
\Im(\bm{X})
\end{array}
\right ]
+
\left[
\begin{array}{c}
\Re(\bm{N})\\
\Im(\bm{N})
\end{array}
\right ],
\end{equation}
where $\Re(\cdot)$ and $\Im(\cdot)$ denote the real and imaginary parts, respectively. The real and imaginary parts are then concatenated together and vectorized as
\begin{equation}\label{net2}
\tilde{\bm{Y}}=\text{vec}([\Re(\bm{Y})^{T}, \Im(\bm{Y})^{T}]^T) \in \mathbb{R}^{2LM \times 1},
\end{equation}
\begin{equation}\label{net3}
\tilde{\bm{S}} =\left[[\Re(\bm{S}), -\Im(\bm{S})]^T,[\Im(\bm{S}), \Re(\bm{S})]^T \right]^{T} \otimes \bm{I}_M  \in \mathbb{R}^{2LM \times 2NQM},
\end{equation}
\begin{equation}\label{net4}
\tilde{\bm{X}}=\text{vec}([\Re(\bm{X})^{T}, \Im(\bm{X})^{T}]^{T}) \in \mathbb{R}^{2NQM \times 1},
\end{equation}
\begin{equation}\label{net5}
\tilde{\bm{N}}=\text{vec}([\Re(\bm{N})^{T}, \Im(\bm{N})^{T}]^{T}) \in \mathbb{R}^{2LM \times 1},
\end{equation}
where $\text{vec}(\cdot)$ is the vectorize operation that flattens a matrix into a vector in the order of columns, and $\otimes$ is the Kronecker product operator. Consequently, (\ref{sysm7}) can be rewritten as
\begin{equation}\label{net6}
\tilde{\bm{Y}}=\tilde{\bm{S}}\tilde{\bm{X}}+\tilde{\bm{N}}.
\end{equation}

According to the recursive formula in (\ref{amp1})-(\ref{amp2}), the input to the DL-mAMPnet is chosen to be the the received signal, the estimated signal, and the residual, which are initialized as $\tilde{\bm{X}}_0=\bm{0}$ and $\tilde{\bm{R}}_0=\tilde{\bm{Y}}$. Meanwhile, unlike the existing AMP-inspired network that uses $\tilde{\bm{X}}$ \cite{27}, we adopt $\bm{\alpha} = [\alpha_1^1, \cdots, \alpha_1^Q, \alpha_2^1, \cdots, \alpha_N^Q]^{T} \in \{0,1 \}^{NQ\times1}$ as the output of DL-mAMPnet, such that $\alpha_n^q$ can be directly obtained once DL-mAMPnet is well-trained.

\begin{figure}[!ht]
  \centering
  \includegraphics[width=7cm]{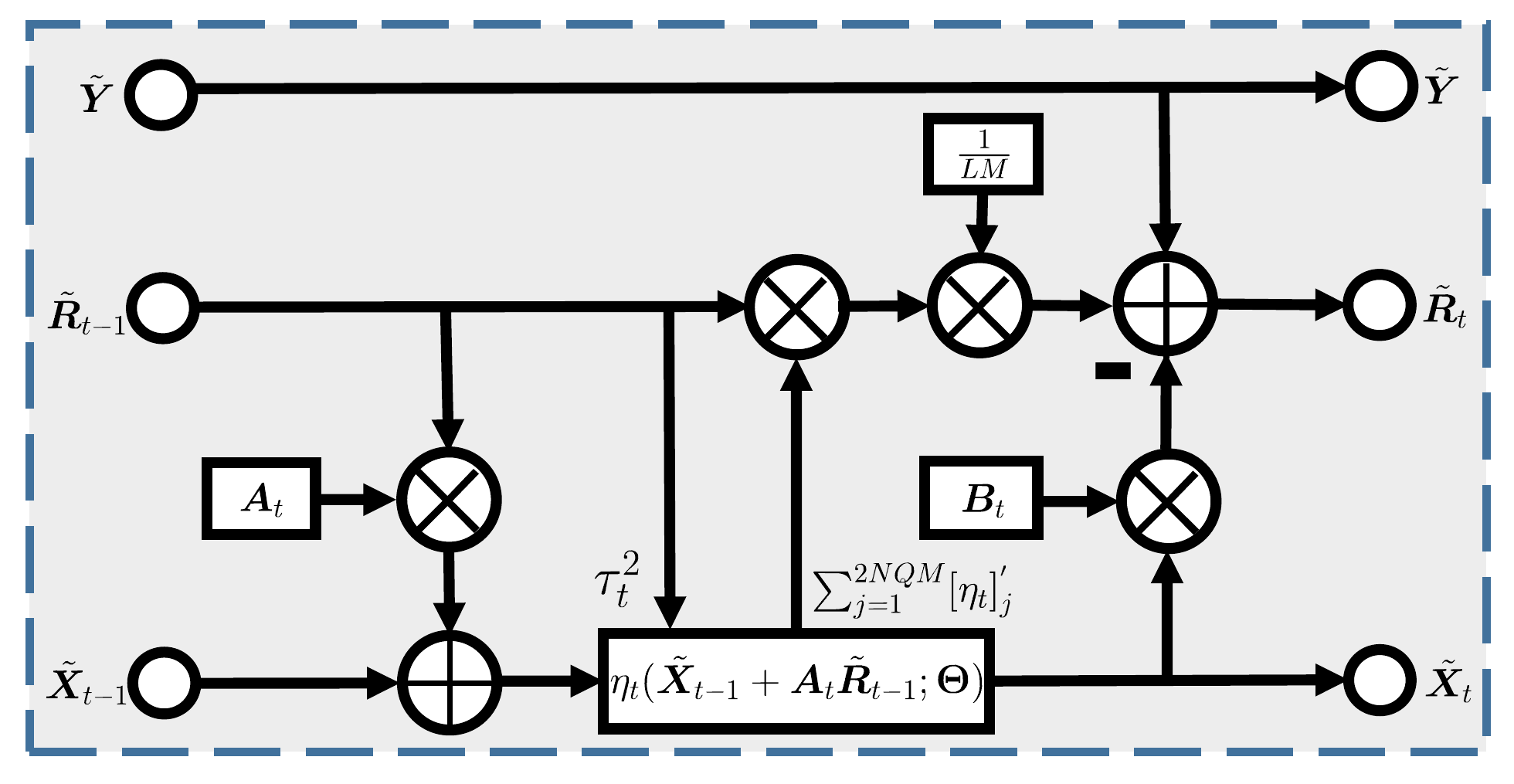}\\
  \caption{Detailed structure of the $t$th AMP layer.}\label{fig3}
\end{figure}

\subsection{AMP Layer}
Since each layer has the same structure, we focus on the  $t$th AMP layer of the DL-mAMPnet, of which the detailed structure is illustrated in Fig.~\ref{fig3}. Define the input as $\tilde{\bm{X}}_{t-1}$, $\tilde{\bm{R}}_{t-1}$ and the output as $\tilde{\bm{X}}_{t}$, $\tilde{\bm{R}}_{t}$, the $t$th AMP layer proceeds as follows
\begin{equation}\label{net7}
\tilde{\bm{X}}_{t}=\eta_{t}(\tilde{\bm{X}}_{t-1}+\bm{B}_{t}\tilde{\bm{R}}_{t-1}; \bm{\Theta}_t),
\end{equation}
\begin{equation}\label{net8}
\tilde{\bm{R}}_{t}=\tilde{\bm{Y}}-\bm{A}_t\tilde{\bm{X}}_{t}+ \frac{\tilde{\bm{R}}_{t-1}}{LM}\sum_{j=1}^{2NQM} [\eta_{t}(\tilde{\bm{X}}_{t-1}+\bm{B}_{t}\tilde{\bm{R}}_{t-1};\bm{\Theta}_t)]^{'}_{j},
\end{equation}
where $\bm{A}_{t}$ and $\bm{B}_{t}$ are trainable matrices that acts as the matched filter and $\bm{\Theta}_t = \{\bm{\theta}_{t,1}, \bm{\theta}_{t,2}\}$ is the trainable parameter set of $\eta_{t}(\cdot)$.

It should be mentioned that the denoiser in (\ref{amp15})-(\ref{amp19}) cannot be applied in the AMP layer, as the complex-to-real transformation and vectorization in (\ref{net2})-(\ref{net6}) have changed the dimension and distribution of the corresponding matrices. Following the same derivation in Appendix A but considering $\tilde{\bm{X}}$ as a real-valued Bernoulli Gaussian variable and changing the dimension, $\eta_{t}(\cdot)$ in (\ref{amp15}) can be expressed as
\begin{align}\label{net9}
&[\eta_{t}(\tilde{\bm{Z}})]_j = \frac{\beta \tilde{\bm{Z}}_j}{(\beta+\tau_t^2)\left(1+\frac{Q-\epsilon}{\epsilon}\exp(\log(1+\frac{\beta}{\tau_t^2})^{1/2}-\frac{\tilde{\bm{Z}}_j^2\beta}{2(\beta+\tau_t^2)\tau_t^2})\right)}, \nonumber \\
& = \frac{\tilde{\bm{Z}}_j}{(1+\frac{\tau_t^2}{\beta})\left(1+\sqrt{1+\frac{\beta}{\tau_t^2}}\exp(\log(\frac{Q-\epsilon}{\epsilon})-\frac{\tilde{\bm{Z}}_j^2}{2(\tau_t^2+\tau_t^4/\beta)})\right)},
\end{align}
where $\tilde{\bm{Z}}_j$ is the $j$th element of $\tilde{\bm{Z}}$.

As discussed in Section II-D, $\eta_{t}(\cdot)$ exploits an i.i.d. assumption that fails to effectively explore the correlated sparsity pattern. To tackle this issue, we replace $\log(\frac{Q-\epsilon}{\epsilon})$ with a
trainable parameter $\bm{\theta_{t,1}} = [ \theta_{t,1,1}, \cdots, \theta_{t,1,2NQM} ]^{T}\in \mathbb{R}^{2NQM \times 1}$, such that the correlation among entries of $\tilde{\bm{X}}$ can be learned and approximated. Meanwhile, to circumvent the need for the prior information of the large-scale fading, we introduce a trainable parameter $\bm{\theta_{t,2}} = [ \theta_{t,2,1}, \cdots, \theta_{t,2,2NQM} ]^{T} \in \mathbb{R}^{2NQM \times 1}$ and
substitute it for $\beta$ in (\ref{net9}). The trainable $\eta_{t}(\cdot)$ can then be defined as
\begin{equation}\label{net10}
[\eta_{t}(\tilde{\bm{Z}})]_j = \frac{\tilde{\bm{Z}}_j}{(1+\frac{\tau_t^2}{\theta_{t,2,j}})\left(1+\sqrt{1+\frac{\theta_{t,2,j}}{\tau_t^2}}\exp(\theta_{t,1,j}-\frac{\tilde{\bm{Z}}_j^2}{2(\tau_t^2+\tau_t^4/\theta_{t,2,j})})\right)}.
\end{equation}

The derivative of $\eta_{t}(\cdot)$ is thus be given by
\begin{equation}\label{net11}
[\eta_{t}(\tilde{\bm{Z}})]_j^{'}  = \frac{[\eta_{t}(\tilde{\bm{Z}})]_j}{\partial \tilde{\bm{Z}}_j }  =\frac{1+\sqrt{1+\frac{\theta_{t,2,j}}{\tau_t^2}}\exp(\theta_{t,1,j}-\frac{\tilde{\bm{Z}}_j^2}{2(\tau_t^2+\tau_t^4/\theta_{t,2,j})})(1+\frac{\tilde{\bm{Z}}_j^2}{(\tau_t^2+\tau_t^4/\theta_{t,2,j})})}{(1+\frac{\tau_t^2}{\theta_{t,2,j}})\left(1+\sqrt{1+\frac{\theta_{t,2,j}}{\tau_t^2}}\exp(\theta_{t,1,j}-\frac{\tilde{\bm{Z}}_j^2}{2(\tau_t^2+\tau_t^4/\theta_{t,2,j})})\right)^2}.
\end{equation}

Note that to evade the computation of the expectation involved in $\tau^2$, this paper adopts an empirical result where $\tau^2$ is estimated by the standard deviation of the corrupted noise in $\tilde{\bm{Z}}$, i.e., $\tau_t^2=||\tilde{\bm{R}}_t||_2 /\sqrt{2LM}$ \cite{27}.

\emph{Remark 1:}
It is worth noting that the denoiser derived in (\ref{amp15}) operates in a section-wise manner, i.e., acts on $Q$ rows of each $\bm{X}_n$, while the $\eta_{t}(\cdot)$ in the AMP layer operates row-by-row on $\bm{X}$. Although the section-wise manner may exploit the correlations better than the row-wise manner, it is quite challenging to be implemented in DNNs. This is because to realize such section-wise manner, we have to either construct $N$ sublayers or impose $N$ iterations in each AMP layer. The former will heavily expand the network size and trainable parameters, reducing the scalability and stunting the training process of the DL-mAMPnet. The latter will greatly increase the computational complexity of the DL-mAMPnet and negate the ``deep unfolding" advantage. It should also be noted that although the AMP layer can explore the correlated sparsity pattern with the help of trainable parameters, the device-level sparsity constraint in (\ref{sysm8}) is not surely guaranteed. Motivated by this consideration, we propose a felicitous method in the refinement module that utilizes the Maxpool-MaxUnpool operation to ensure device-level sparsity, as detailed in the subsection below.

\begin{figure*}[!ht]
  \centering
  \includegraphics[width=13cm]{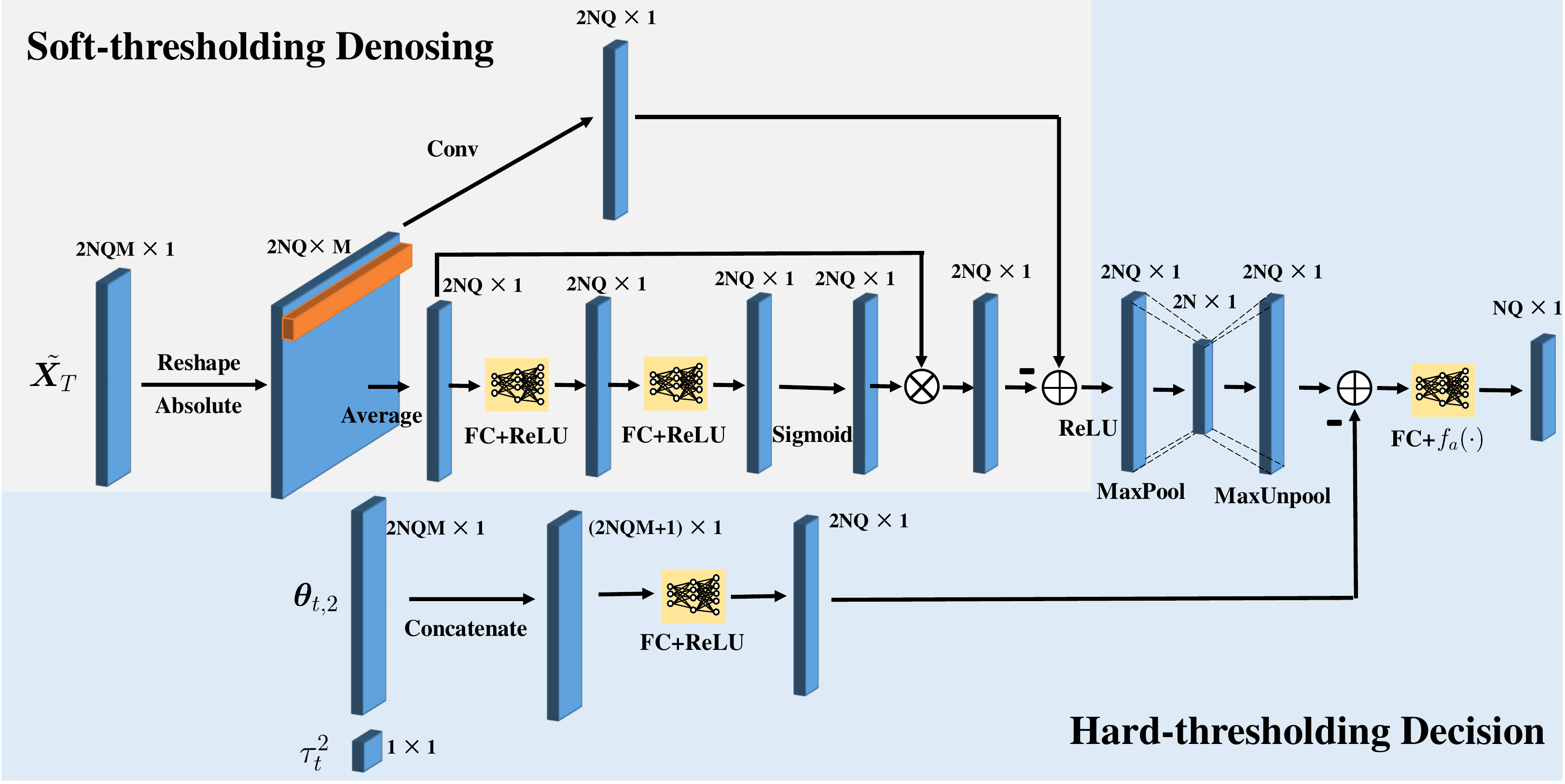}\\
  \caption{Detailed architecture of the proposed refinement module.}\label{fig4}
\end{figure*}

\subsection{Refinement Module}
The refinement module should be capable of ensuring the device-level sparsity while extracting $a_n^q$ from $\tilde{\bm{X}}_{T}$ without explicit channel state information (CSI). To fulfil these functionalities, two components are  integrated in the refinement module, namely the \emph{soft-thresholding denoising component} and the \emph{hard-thresholding decision component}. The soft-thresholding denoising component is intended to further denoise $\tilde{\bm{X}}_{T}$ by exploiting the hierarchical sparse structure. The hard-thresholding decision component is aimed at implementing the threshold-based strategy in (\ref{rm5})-(\ref{rm7}). The detailed structure of the refinement module is presented in Fig.~\ref{fig4} and elaborated as follows.

\textbf{Soft-Thresholding Denoising:}
As shown in Fig.~\ref{fig1}, the two-level sparsity exhibits a unique spatial structure that has not been utilized in the AMP layers. Here, the soft-thresholding denoising aims to distill $\tilde{\bm{X}}_{T}$ using such spatial feature, enhancing useful information while removing noise information. To do this, we first de-vectorize $\tilde{\bm{X}}_{T}$ and take the absolute value as
\begin{equation}\label{rm1}
\overline{\bm{X}} = |\text{Vec}^{-1}(\tilde{\bm{X}}_{T})| =[|\Re(\bm{X})^{T}|, |\Im(\bm{X})^{T}|]^{T} \in {\mathbb{R}^{+}}^{2NQ \times M}.
\end{equation}
Then, a convolutional layer with $1 \times M$ kernel size is applied to $\overline{\bm{X}}$ to combine the information from all $M$ antennas and extract a coarse estimation of $a_n^{q}$. This arrangement is motivated by the fact that all $M$ elements in each row of $\overline{\bm{X}}$ share the same $a_n^{q}$, as observed from (\ref{sysm6}) and Fig.~\ref{fig1}. The coarse estimation can be expressed as $f_{\theta_c}(\overline{\bm{X}})$, where $f_{\theta_c}(\cdot)$ is the function expression of the convolutional layer with parameter $\theta_c$. After that, an average pooling with $1 \times M$ kernel size is applied to $\overline{\bm{X}}$ to get a 1-D average vector over $M$ antennas. The 1-D vector $\bm{\iota}=\frac{1}{M}\sum_{m=1}^{M} \overline{\bm{X}}_{:,m}$ is forwarded into a two-layer fully-connected (FC) network to obtain a scaling parameter, such that the inner features of the average value among the $2NQ$ rows of $\overline{\bm{X}} $ can be learned.  The scaling parameter is then scaled to the range of $(0, 1)$ using a sigmoid function, which can be written as follows
\begin{equation}\label{rm2}
\bm{\vartheta} = \frac{1}{1+e^{-f_{\theta_{FC_1}}(\bm{\iota})}},
\end{equation}
where $\bm{\vartheta}$ is the scaling vector and $f_{\theta_{FC_1}}(\cdot)$ is the function expression of the two-layer FC network with parameter $\theta_{FC_1}$. Next, $\bm{\vartheta}$ is multiplied by $\bm{\iota}$ to get the threshold as
\begin{equation}\label{rm3}
\bm{\kappa}_{ST} = \bm{\vartheta} \odot \bm{\iota},
\end{equation}
where $\odot$ is the Hadamard product operator. This operation is inspired by the fact that the threshold for soft thresholding must be positive and not too large \cite{28}. If the threshold is larger than the largest value of $f_{\theta_c}(\overline{\bm{X}})$, then the output of soft thresholding will all be zeros, and thus the useful information will be removed. Finally,  the obtained threshold $\bm{\kappa}_{ST}$ is subtracted by $f_{\theta_c}(\overline{\bm{X}})$ and fed into a ReLU activation function as
\begin{equation}\label{rm4}
\bm{o} = \max(0,f_{\theta_c}(\overline{\bm{X}})-\bm{\kappa}_{ST}),
\end{equation}
where $\bm{o}$ denotes the output of the soft-thresholding denoising component. We can observe from (\ref{rm4}) that by keeping $\bm{\kappa}_{ST}$ in a reasonable range, the useful information can be preserved while the noise information is eliminated. It is worth noting that, rather than being manually set by experts, such a threshold can be learned automatically in the proposed soft-thresholding denoising component, removing the need for the expertise of signal processing and the statistical characteristic of $\overline{\bm{X}}$.

\begin{figure}[!ht]
  \centering
  \includegraphics[width=6cm]{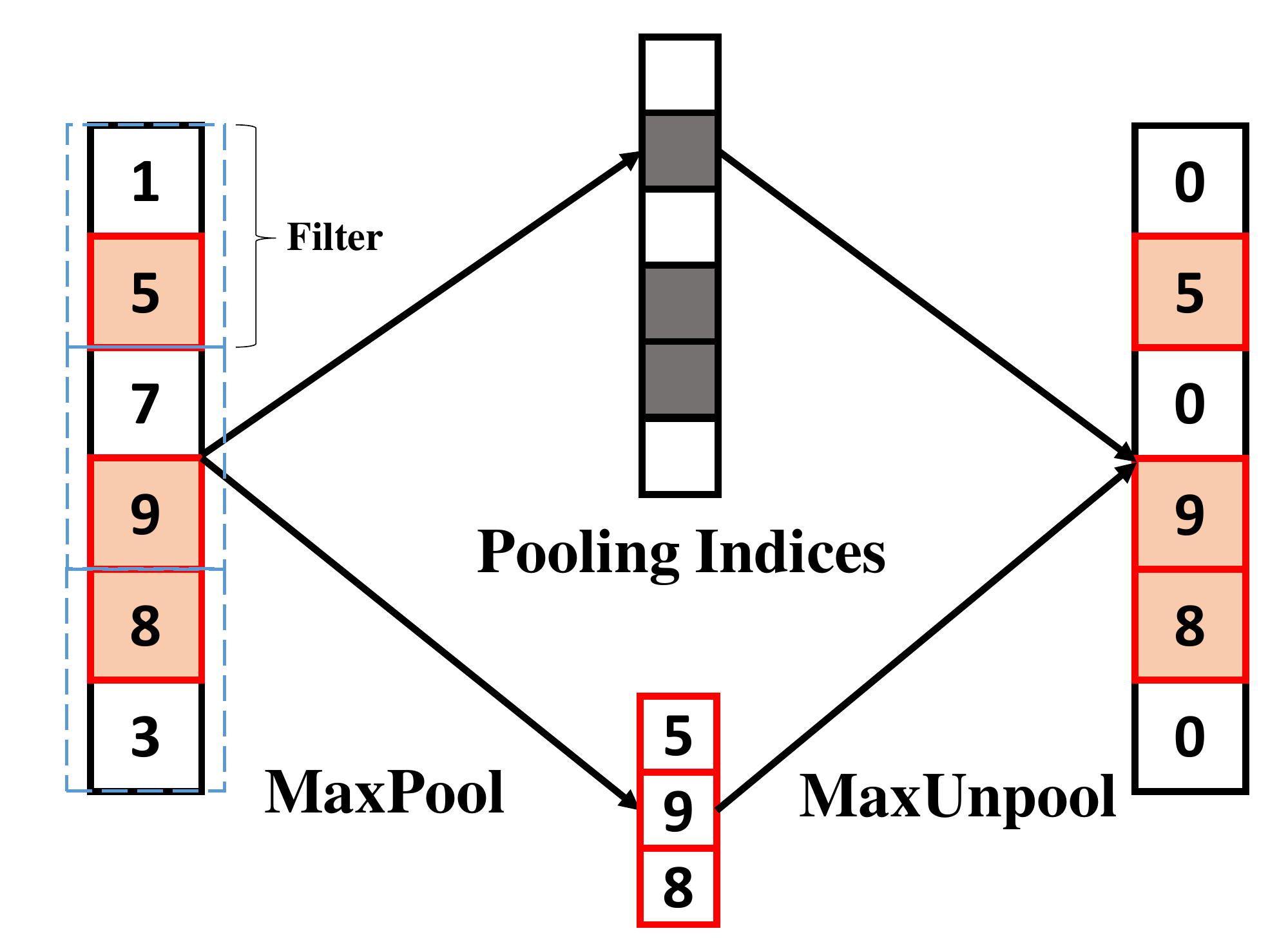}\\
  \caption{Illustration of the MaxPool-MaxUnpool process.}\label{fig5}
\end{figure}

\textbf{Hard-thresholding Decision:}
It is challenging to directly implement the threshold-based strategy in DNNs, as (\ref{rm5}) is non-differentiable and will stunt the backpropagation process. To tackle this issue, the hard-thresholding decision component elegantly uses the \emph{Maxpool} and \emph{MaxUnpool} procedures to ensure the device-level sparsity. \emph{Maxpool} is a down-sampling technique that uses a max filter to non-overlapping subregions of the initial input \cite{29}. For each region represented by the filter,  we will take the max of that region and create a new output matrix where each element is the max of a region in the original input. \emph{Maxunpool}, in contrast, expands the output of the maxpool operation to its original size by upsampling and padding with zeros. Except for the maximum position, all the rest elements in the unpooled matrix are supplemented with $0$.

For an intuitive explanation, we illustrate the process of \emph{Maxpool} and \emph{MaxUnpool} in Fig.~\ref{fig5}. It can be observed from Fig.~\ref{fig5} that in each filter, except for the largest value that remains unchanged, all the rest elements become 0. Such manipulation perfectly executes the element selection operation in (\ref{amp5}). By setting the filter size as $Q \times 1$, we enforce that at most one non-zero row exists in the $Q$ rows of $\bm{X}_n$, and therefore the device-level sparsity constraint in (\ref{c2}) can be guaranteed. It should also be mentioned that the pooling procedure is only a module that alters the dimension size during the deep learning process, which has no parameters and thus has no impact on network training.

After guaranteeing the device-level sparsity, the onus shifts to performing the threshold-based decisive operation in (\ref{rm6}), i.e.,
determining the binary sequence $\bm{\alpha}$ by comparing the threshold $\kappa_{n}^{i^{*}_n}$ with the matrix obtained from the maxpool-maxunpool procedure $\text{Mp}(\text{Mup}(\bm{o}))$. However, some issues exist when determining $\bm{\alpha}$. The first issue is that the threshold in (\ref{rm7}) may not be precise sufficiently because it is derived under an mismatched i.i.d. assumption. To tackle this issue, we look afresh at (\ref{rm7}) and find that the threshold is a function of $\beta$ and $\tau$. Since $\beta$ has been represented by $\bm{\theta}_2$ in (\ref{net10}), we concatenate $\bm{\theta}_{T,2} $ and $\tau_T^2$ outputted from the last AMP layer and feed it into an FC layer with ReLU activation function to learn the accurate threshold, which is denoted by
\begin{equation}\label{rm8}
\bm{\kappa}_{HT} = \max(0,f_{\theta_{FC2}}(\bm{\theta}_{T,2},\tau_T^2)),
\end{equation}
where $f_{\theta_{FC2}}$ is the function expression of the FC network with parameter $\theta_{FC2}$.

Then, the learned threshold $\bm{\kappa}_{HT}$ is subtracted by $\text{Mp}(\text{Mup}(\bm{o}))$ and forwarded into an FC layer with parameter $\theta_{FC3}$ to fulfil the threshold-based decisive operation. The FC layer here has two functionalities: compressing the dimension from $2NQ\times1$ to $NQ\times1$ and converting the $\bm{\kappa}_{HT}$-$\text{Mp}(\text{Mup}(\bm{o}))$ difference into a binary sequence. Mathematically, the optimal function for threshold-based binary decision is the signum function denoted as
\begin{equation}
\label{rm9}
\text{sng}(x)=\left\{
\begin{aligned}
1 & , & x >0; \\
0 & , & x \le 0.
\end{aligned}
\right.
\end{equation}
However, since $\text{sng}(x)$ is non-differentiable, it cannot be used in DNN, necessitating the development of a substitute function.

\begin{figure}[!ht]
  \centering
  \includegraphics[width=7cm]{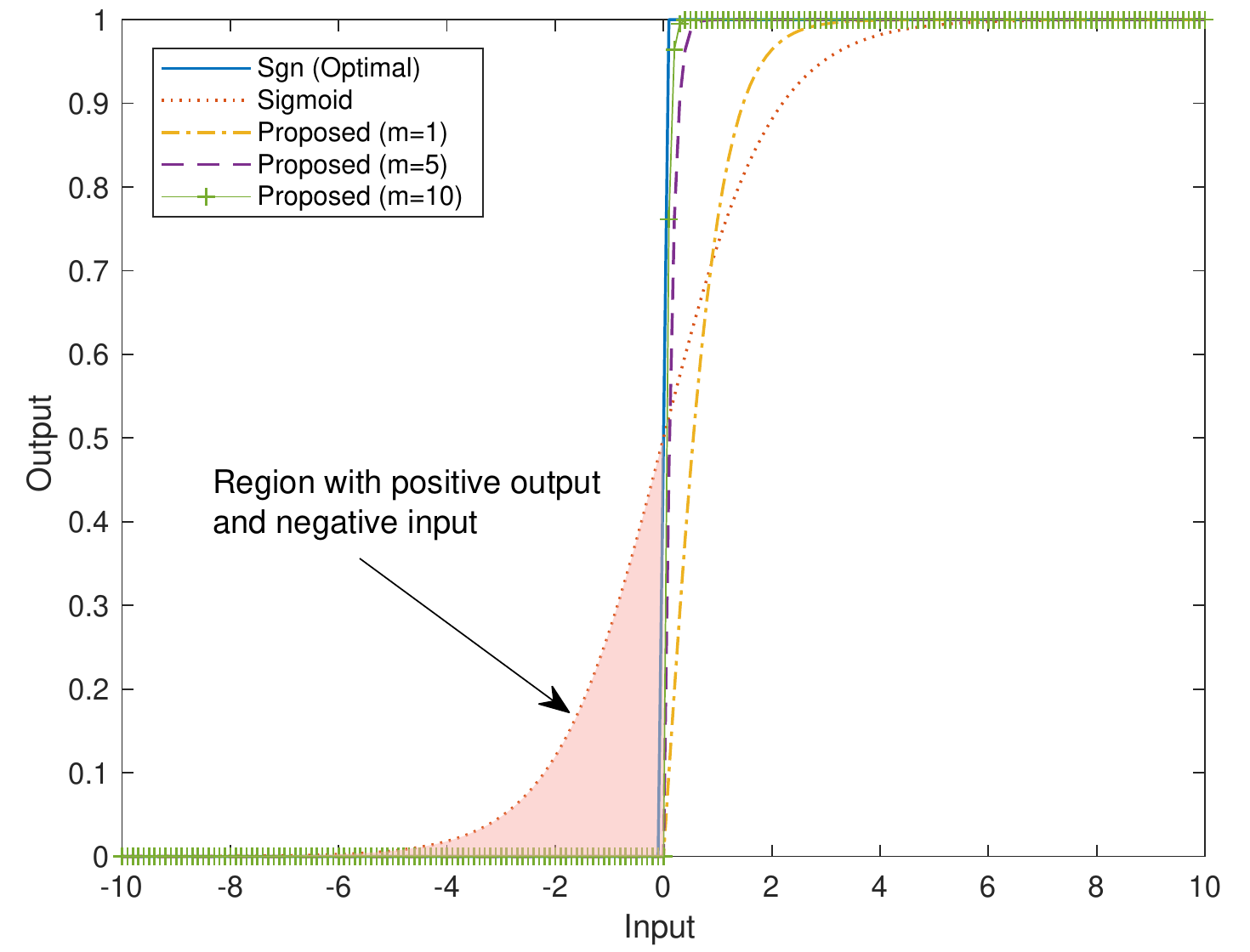}\\
  \caption{The curves of the optimal signum, sigmoid, and hard-thresholding decision functions.}\label{fig6}
\end{figure}

When it comes to DL-based binary decisions, the sigmoid function is a popular choice and has been widely used in the literature \cite{30}, as it can map the input to the interval within [0, 1]. The sigmoid function, nevertheless, is still inapplicable to the hard-thresholding decision module. The reasons are as follows: (i) The sigmoid function returns a continuous value between 0 and 1, implying that a threshold is further required to distinguish the outputted value as 0 or 1. However, it is usually non-trivial to design an appropriate threshold; (ii) According to (\ref{rm6}), the output of the threshold-based decision should be strictly 0 with negative input. However, as shown in Fig.~\ref{fig6}, there is a region where the output is still positive with negative input in the sigmoid function, which may introduce additional errors. To solve the above issues, we devise a novel hard-thresholding decision function, whose core idea is to cascade the ReLU function with tahn function and introduce a multiplier $\varrho$ to approximate the cascaded function as a signum function. The proposed hard-thresholding decision function is given by
\begin{equation}\label{rm10}
f_{\varrho}(x)=\max(0,\frac{e^{\varrho x}-e^{-\varrho x}}{e^{\varrho x}+e^{-\varrho x}}).
\end{equation}
By cascading the ReLU function with tahn function, we not only ensure that the output of the threshold-based decision is strictly 0 with negative input, but also guarantee the output with positive input approximates to 1 with the increment of $\varrho$. The optimal signum, sigmoid, and hard-thresholding decision functions are plotted in Fig.~\ref{fig6}. The figure shows that with the increase of $\varrho$, $f_{\varrho}(\cdot)$ gradually approximates to $\text{sng}(x)$, validating the rationality of the proposed hard-thresholding decision function.

\emph{Remark 2:}
Although we restrict the application of the hard-thresholding decision component to the non-coherent transmission in mMTC, the proposed component can be used in any other scenarios where the signal has a special sparsity structure, such as the spatial modulation system. Meanwhile, the devised hard-thresholding decision function can also be used in any bit-level detector. That is, the hard-thresholding decision component is a plug-and-play module with a wide range of applications.

\section{The Implementation of DL-mAMPnet}

\subsection{Parameter Initialization}
In deep learning, parameter initialization plays a critical role in speeding up convergence and achieving lower error rates. Choosing proper initialization values is especially important for the proposed DL-mAMPnet, as the DL-mAMPnet is built on the AMP algorithm and thus should preserve some essential features to ensure performance and interpretability. There are mainly three items needed to be considered for parametrization: the trainable matrices $\bm{A}_t$ and $\bm{B}_t$, the denoiser parameter set $\bm{\Theta}_t$, and the refinement module parameters $\bm{\theta_{RM}}=\{\theta_{FC_1}, \theta_{FC_2}, \theta_{FC_3}, \theta_{C}\}$.

\subsubsection{Initializing $\bm{A}_t$ and $\bm{B}_t$}
It can be observed from (\ref{net7})-(\ref{net8}) that the DL-mAMPnet implements a generalization of the AMP algorithm in (\ref{amp1})-(\ref{amp2}), wherein the matched filters $(\bm{S}, \bm{S}_n^H)$ manifest as $(\bm{A}_t, \bm{B}_{t})$ at iteration $t$. However, such generalization does not enforce $\bm{B}_{t}=\bm{A}_{t}^H$ and thus may not preserve the independent-Gaussian nature of the denoiser input (\ref{amp6}). According to the analysis in \cite{27}, the desired nature maintains when $\bm{A}_t=\upsilon_t \bm{S}$ with $\upsilon_t>0$. Therefore, $\bm{A}_t$ is parameterized as $\upsilon_t\bm{S}$ and (\ref{net7})-(\ref{net8}) can be rewritten as
\begin{equation}\label{pi1}
\tilde{\bm{X}}_{t}=\upsilon_t\eta_{t}(\tilde{\bm{X}}_{t-1}+\bm{B}_{t}\tilde{\bm{R}}_{t-1}; \bm{\Theta}_t),
\end{equation}
\begin{equation}\label{pi2}
\tilde{\bm{R}}_{t}=\tilde{\bm{Y}}-\bm{S}\tilde{\bm{X}}_{t}+ \frac{\upsilon_t\tilde{\bm{R}}_{t-1}}{LM}\sum_{j=1}^{2NQM} [\eta_{t}(\tilde{\bm{X}}_{t-1}+\bm{B}_{t}\tilde{\bm{R}}_{t-1}; \bm{\Theta}_t)]^{'}_{j},
\end{equation}
the derivation of which can be found in \cite{27} and is omitted here for brevity. In this paper, we initialize $\bm{B}_t = \tilde{\bm{S}}^{T}$ and $\upsilon_t=1$, since such initialization can greatly expedite the convergence of the training process \cite{27}.

\subsubsection{Initializing $\bm{\Theta}_t$}
For $\bm{\theta}_1$, we initialize each element as $\log(\frac{Q-\epsilon}{\epsilon})$, i.e., initialize that each pilot sequence has the same active probability. This is because we have no prior information about the device activity and the transmitted pilot sequence index.  By adopting such a uniform initialization, the initial $\bm{\theta}_1$ will have the minimum Euclidean distance from the actual value. For example, consider a device with a 2-bit message and active indicator $\{1,0,0,0\}$. If we start with a mismatched one-hot vector, then the Euclidean distance will be $\sqrt{2}$. If we initialize $\bm{\alpha}_n$ as $\{\frac{1}{4},\frac{1}{4},\frac{1}{4},\frac{1}{4}\}$, then the  Euclidean distance will be $\sqrt{\frac{3}{4}}$. Therefore, the uniform initialization can accelerate the convergence as a shorter Euclidean distance may lead to faster convergence.

The initial value of $\bm{\theta}_2$ can be computed from the received signal strength. Recall that each pilot sequence has a unit norm and $\bm{h}_n \sim \mathcal{CN}(\bm{0},\beta_n\bm{I}_M)$, each element of the initial
$\bm{\theta}_2$ is roughly given by $||\tilde{\bm{Y}}||_2^2 /\sqrt{2K}$.

\subsubsection{Initializing $\bm{\theta_{RM}}$}
For all parameters in the refinement module, we adopt the He initialization \cite{31} as it has been mathematically proved to be the best weight initialization strategy for the ReLU activation function \cite{32}.

\subsection{Parameter Training}
\subsubsection{Training Algorithm}
Aside from the network structure and parameter initialization, the training algorithm also determines the performance of the DL-mAMPnet. The standard training strategy is the end-to-end training where all the parameters are optimized simultaneously by following the back-propagation rule. However, the end-to-end training is not appropriate for the DL-mAMPnet due to the following reasons: (i) The AMP algorithm aims to provide an estimate $\hat{\bm{X}}(\bm{Y})$ based on $\bm{Y}$ that minimizes the MSE $\mathbb{E}_{\bm{X}\bm{Y}}||\hat{\bm{X}}(\bm{Y})-\bm{X}||_2^{2}$. If the DL-mAMPnet is trained to learn the direct mapping from $\bm{Y}$ to $\bm{\alpha}$, the MSE optimality of the AMP layers may not be achieved; (ii) Even if the AMP layers and the refinement module are trained separately,  the AMP layers can still easily converge to a bad local optimal solution due to overfitting \cite{33}.

For these reasons, we propose a layer-wise training strategy, the idea behind which is to decouple the training of each layer. The details are given in \textbf{Algorithm 1}. There are totally $T+2$ phases in the layer-wise training. In the first phase, we train the learnable parameters of the first AMP layer. Then in the $t$ phase, we train the first $t$ AMP layers with the parameters of the first $t-1$ AMP layers fixed as the parameters learned by the first $t-1$ phases. In the $T+1$ phase, we train the whole network with only the parameters of the refinement module is learnable, while the parameters of the AMP layers are fixed as the parameters learned by the first $T$ phases. Finally, in the last phase, all the parameters are initialized as the parameters learned during  the first $T+1$ phases and then trained jointly.

\renewcommand{\algorithmicrequire}{\textbf{Input:}\unskip}
\renewcommand{\algorithmicensure}{\textbf{Output:}\unskip}

\begin{algorithm}
  \caption{Parameter training of the DL-mAMPnet via layer-wise training strategy}
  \label{alg:bp}
  \begin{algorithmic}
    \REQUIRE Training dataset $D_{AMP}$, $D_{RM}$;
    \ENSURE Trained parameter $\{\upsilon_t, \bm{B}_t, \bm{\Theta}_t\}_{t=1}^{T}$ and $\bm{\theta}_{RM}$;
    \STATE Initialize parameters according to Section IV-B;
    \FOR{$t=1$ to $T$}
      \STATE  Learn $\{\upsilon_t, \bm{B}_t, \bm{\Theta}_t\}_{t}$ with fixed $\{\upsilon_t, \bm{B}_t, \bm{\Theta}_t\}_{t=1}^{t-1}$ based on the loss function (\ref{pt1});
    \ENDFOR
    \STATE Learn $\bm{\theta}_{RM}$ with fixed $\{\upsilon_t, \bm{B}_t, \bm{\Theta}_t\}_{t=1}^{T}$ based on the loss function (\ref{pt2});
    \STATE Re-learn $\{\upsilon_t, \bm{B}_t, \bm{\Theta}_t\}_{t=1}^{T}$ and $\bm{\theta}_{RM}$  based on the loss function (\ref{pt2});
    \RETURN{$\{\upsilon_t, \bm{B}_t, \bm{\Theta}_t\}_{t=1}^{T}$ and $\bm{\theta}_{RM}$.}
  \end{algorithmic}
\end{algorithm}

The training dataset $D_{AMP}$ for the first $T$ phases comprises 100, 000 pairs of $\tilde{\bm{X}}$ and $\tilde{\bm{Y}}$, and the corresponding loss function is the MSE loss
\begin{equation}\label{pt1}
\mathcal{L}_t(\tilde{\bm{Y}}) = ||\tilde{\bm{X}}_t(\tilde{\bm{Y}}) - \tilde{\bm{X}}||_{2}^{2},  t = [1,\cdots,T].
\end{equation}
The training dataset $D_{RM}$ for the last 2 phases has 100, 000 pairs of $\bm{\alpha}$ and $\tilde{\bm{Y}}$, and the loss function is the binary cross entropy loss
\begin{equation}\label{pt2}
\mathcal{L}_t(\tilde{\bm{Y}}) = \frac{1}{NQ} \sum_{i=1}^{NQ} \left(\bm{\alpha}(\tilde{\bm{Y}})_i \log \bm{\alpha}_i +  (1-\bm{\alpha}(\tilde{\bm{Y}})_i)\log(1-\bm{\alpha}_i)\right), t = [T+1,T+2].
\end{equation}
The DL-mAMPnet is trained epoch by epoch with the training dataset using the Adam optimizer, while within an epoch, the whole training dataset is shuffled and split into batches with the size of 500.$\footnote{It should be mentioned that the number of epochs and the learning rate are different for each phase, which are empirically determined in Section V.}$

\subsubsection{Training Dataset}
The training dataset is synthetically generated as follows: (i) Generating $\bm{\alpha}_n$: $K$ active devices are randomly selected among $N$ devices. Then, each active device is randomly assigned with a $Q$-dimensional one-hot vector, and each inactive device is assigned with a $Q$-dimensional zero vector; (ii) Generating $\bm{X}_n$: The uplink channel of device $n$, i.e., $\bm{h}_n$, is first generated according to (\ref{sysm1}). Then $\bm{X}_n$ is obtained by multiplying $\bm{h}_n$ and $\bm{\alpha}_n$; (iii) Generating $\bm{Y}$: The pilot sequence $\bm{S}_n$ is generated by sampling from complex Gaussian distribution with zero mean and variance. Given $\bm{X}_n$ and $\bm{S}_n$, $\bm{Y}$ can be directly obtained according to (\ref{sysm6}).

\section{Simulation Results}
In this section, extensive simulations are provided to verify the effectiveness of the proposed algorithm. The setup is as follows unless otherwise stated. We consider a mMTC system with $N = 100$ devices for illustration purpose, although the proposed algorithm can be used for a much larger-scale system. Each device accesses the BS independently with probability $\epsilon=0.1$ at each coherence block. The large-scale fading coefficient for device $n$ is $\beta_n = 128.1-36.7\log_{10}(d_n)$ in dB, where $d_n$ is the distance between device $n$ and the BS that follows a uniform distribution within [0.05, 1] km. The small-scale fading coefficient for each device follows the i.i.d. multivariate complex Gaussian distribution with zero mean and unit variance. The power spectral density of the AWGN at the BS is assumed to be $-169$ dBm/Hz \cite{8} and the bandwidth of the wireless channel is $1$ MHz.

The number of AMP layers in the DL-mAMPnet is set to be $T =4$. The training epochs and learning rate for each training phase are set to be $ \{2,000, 1,500, 1,000, 1,000, 1,500, 5,000\}$ and $\{2 \times 10^{-5}, 2 \times 10^{-5}, 2 \times 10^{-5}, 2 \times 10^{-5}, 1 \times 10^{-5}, 1 \times 10^{-5}\}$.$\footnote{All the parameters are empirically determined using the general workflow, where the training starts with relatively small values and increases the values until the learning performance cannot be further improved.}$  We train the DL-mAMPnet with 80, 000 training samples and test with 20, 000 data samples, which are randomly drawn from $D_{AMP}$ for the first 4 phases and $D_{RM}$ for the last 2 phases. The DL-mAMPnet is trained and tested by on an x86 PC with one Nvidia GeForce GTX 1080 Ti graphics card, and Pytorch 1.1.0 is employed as the backend. The traditional AMP-based algorithm with $T_{AMP}=50$ iterations and the covariance-based method with  $T_{Cov}=50$ iterations~\cite{14} are employed as the benchmark and evaluated on the same dataset. In addition, the SER is adopted as the performance metric: $\text{SER}=\frac{1}{N}\sum_{n=1}^{N} \mathbb{I}(\hat{\bm{\alpha}}_n\neq\bm{\alpha}_n)$, where $\hat{\bm{\alpha}}_n$ and $\bm{\alpha}_n$ denote the estimated pilot sequence activity for device $n$ and its ground truth, respectively.

\subsection{Performance of the DL-mAMPnet}
\begin{figure}[!ht]
  \centering
  \includegraphics[width=3.0in]{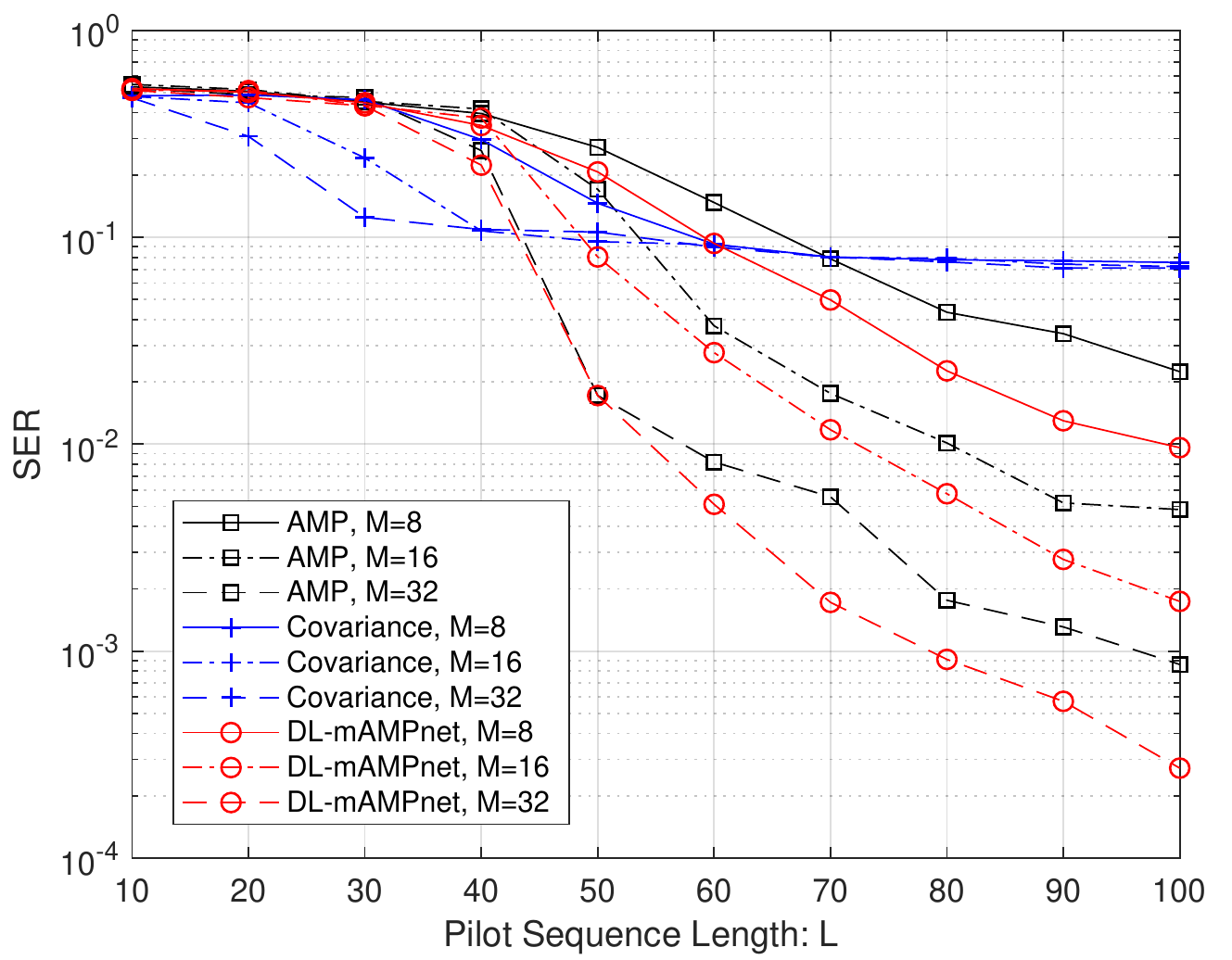}\\
  \caption{SER performance versus the pilot sequence length $L$ for $J=1$ bit.}\label{serl}
\end{figure}

\begin{figure}[!ht]
  \centering
  \includegraphics[width=3.0in]{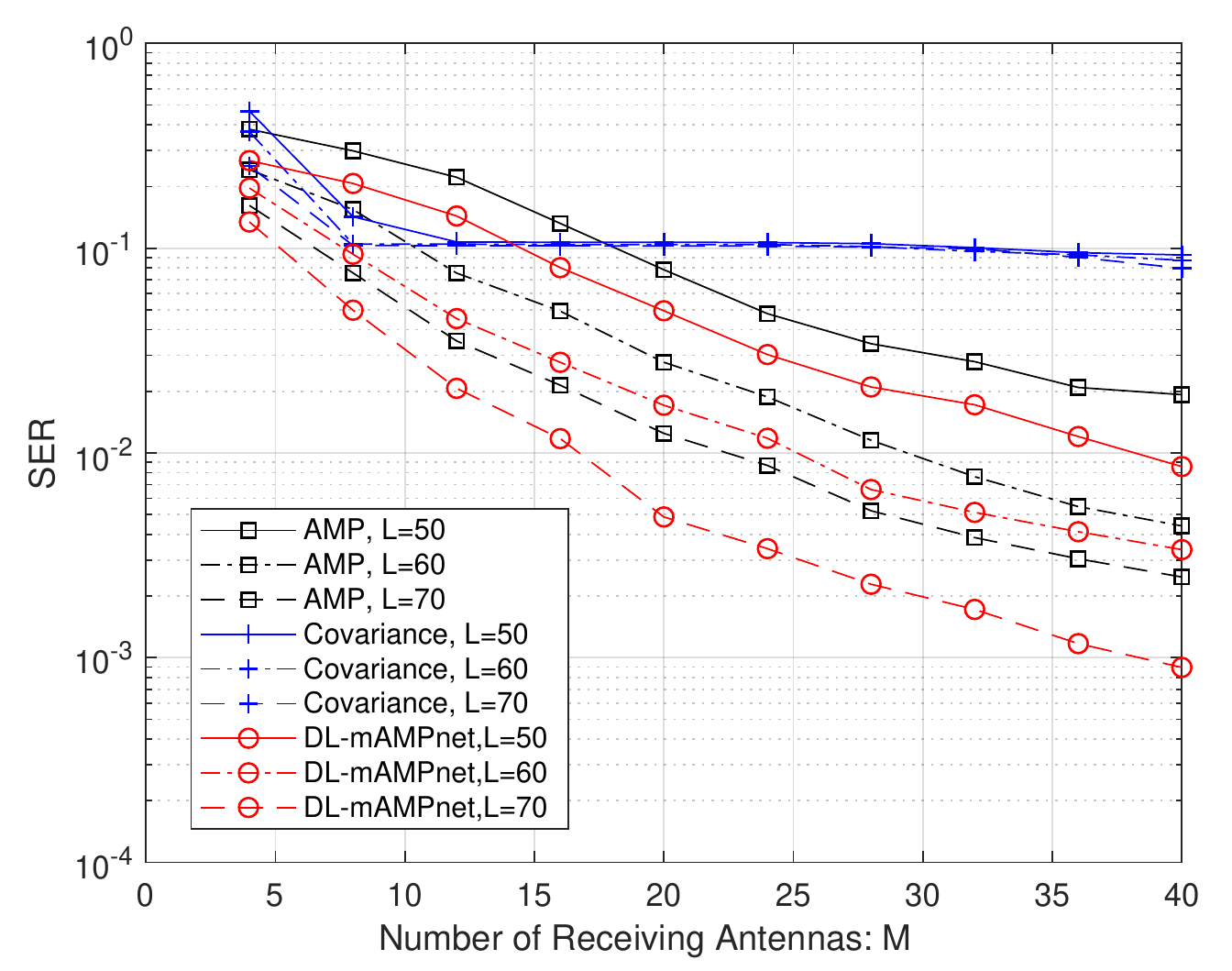}\\
  \caption{SER performance versus the number of receiving antennas $M$ for $J=1$ bit.}\label{serm}
\end{figure}

\begin{figure}[!ht]
  \centering
  \includegraphics[width=3.0in]{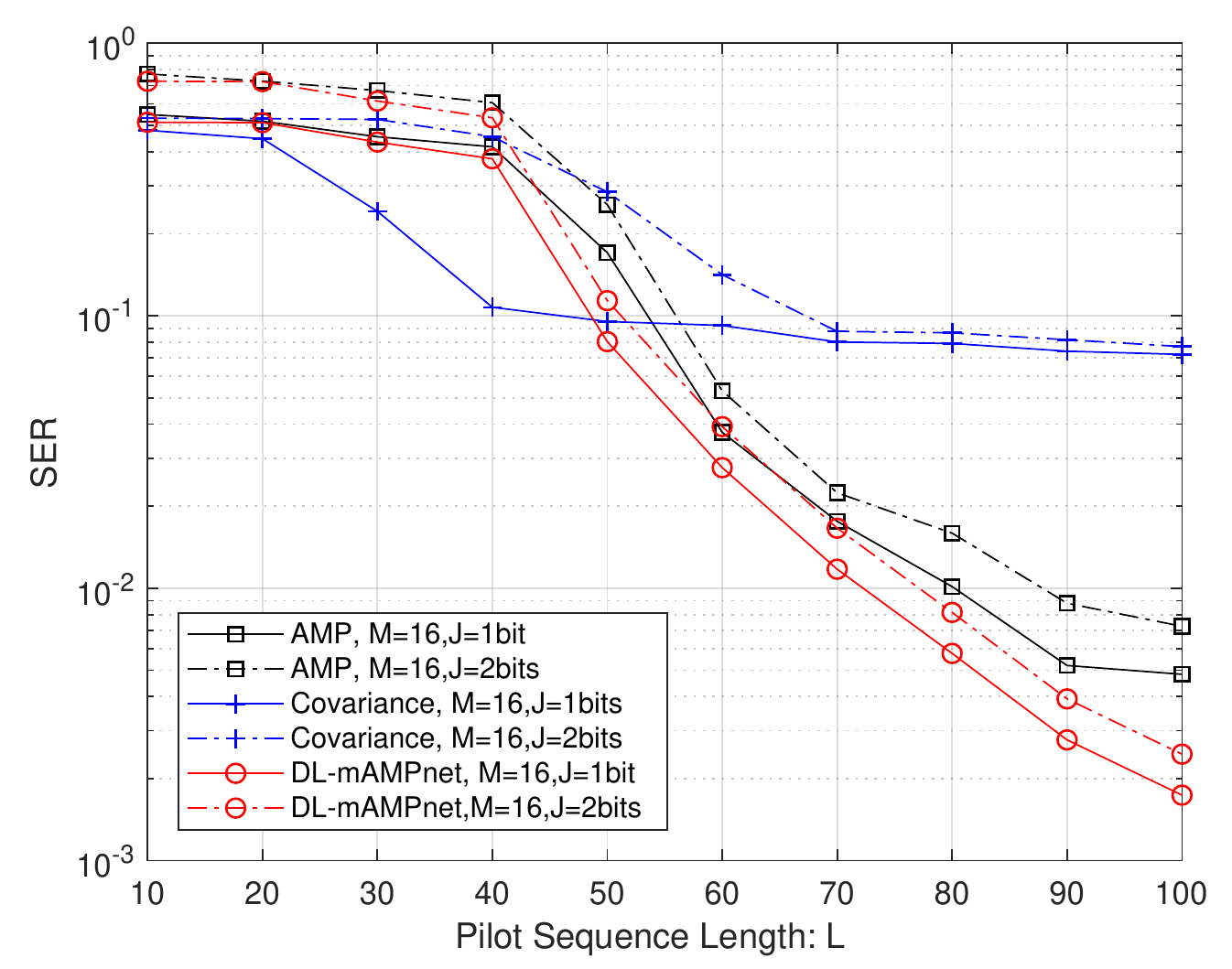}\\
  \caption{SER performance versus the pilot sequence length $L$ with different lengths of transmitted messages $J$. }\label{serq}
\end{figure}

Fig.~\ref{serl} depicts the SER versus $L$ with different values of $M$. It is observed that both the SER of the DL-mAMPnet and AMP-based algorithm decrease as $L$ and $M$ increase. Although the SER of the covariance-based algorithm is lowest when $L$ is small, it becomes saturated when $L$ exceeds some point, e.g., $L = 40$ when $M = 16$. This is mainly due to the suboptimality of the fixed threshold.$\footnote{As observed from (\ref{rm7}), the threshold is variable and related to system parameters such as signal power and receiving antenna numbers, whereas the covariance-based algorithm adopts a fixed threshold. Since there is no concrete method to design such a fixed threshold, we empirically set the threshold of the covariance-based algorithm to be $\beta_n/2$ in this paper.}$  Meanwhile, the proposed DL-mAMPnet notably outperforms the AMP-based algorithm by a large margin. For example, the proposed DL-mAMPnet achieves more than 10 pilot length gain over the AMP-based algorithm when $L$ is larger than 70, which indicates that the proposed DL-mAMPnet can reduce the required pilot sequence length, lowering the difficulty of pilot design and adapting to fast-changing channels. Moreover, although for any $M$, the SERs of both the DL-mAMPnet and AMP-based algorithm decrease over $L$, the reduction is faster when $M$ is $32$ as compared to that when $M$ is $8$, which shows that increasing the number of receiving antennas can further reduce the required pilot sequence length.

Fig.~\ref{serm} shows the SER versus $M$ for various values of $L$. We observe that for the DL-mAMPnet and AMP-based algorithm, the SER drops effectively as $M$ increases, whereas for the covariance-based algorithm, there are error floors in the SER. Moreover, the DL-mAMPnet needs fewer receiving antennas to achieve the same performance as the AMP-based algorithm, implying that the proposed DL-mAMPnet can reduce demand for receiving antennas, resulting in lower deployment cost and energy consumption.

Fig.~\ref{serq} plots the SER versus $L$, with 2 different lengths of transmitted messages, i.e., $J =1$ bit and $J = 2$ bits. The number of receiving antennas is $M=16$. It can be seen that the SERs of all three algorithms increase as the length of transmitted messages increases, which implies that the performance of both algorithms deteriorates when more messages are transmitted. An important point is that as the message length increases, the performance gap between the proposed DL-mAMPnet and the other two algorithms increases, indicating the potential of the DL-mAMPnet to handle long packet size.

\subsection{Visualization of the DL-mAMPnet}

\begin{figure}
\centering
\subfigure[]{
\begin{minipage}[b]{0.135\textwidth}
\includegraphics[width=1\textwidth]{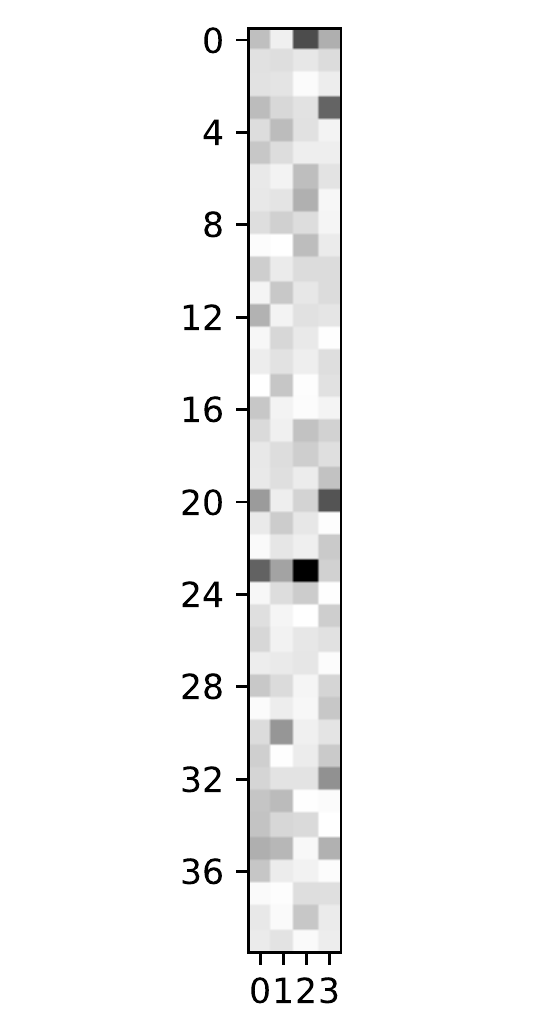}
\end{minipage}
}
\subfigure[]{
\begin{minipage}[b]{0.138\textwidth}
\includegraphics[width=1\textwidth]{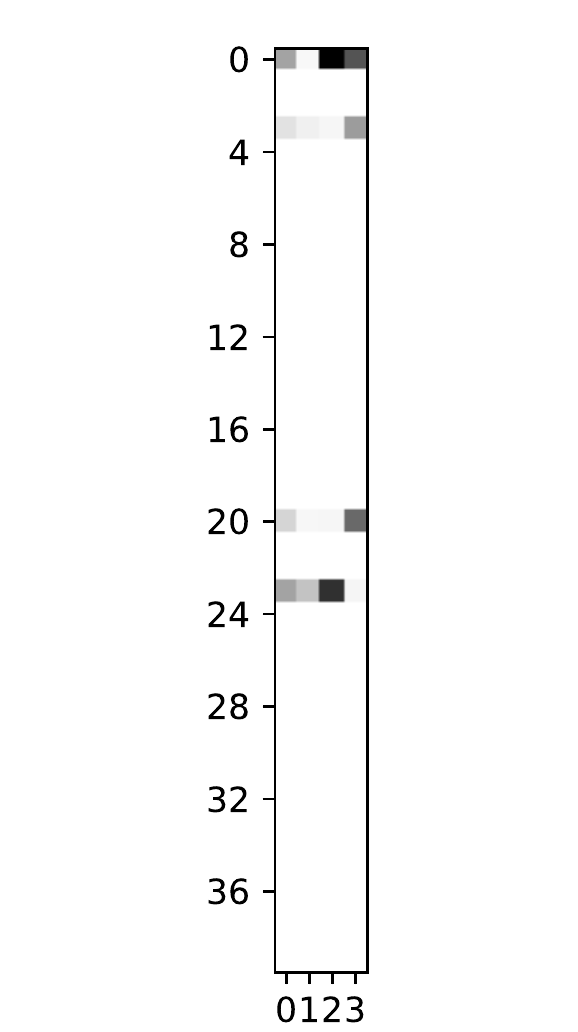}
\end{minipage}
}
\subfigure[]{
\begin{minipage}[b]{0.135\textwidth}
\includegraphics[width=1\textwidth]{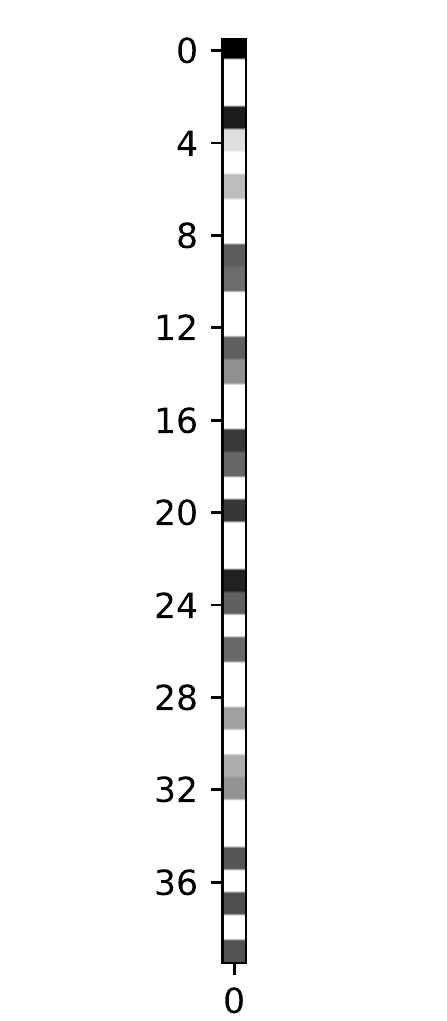}
\end{minipage}
}
\subfigure[]{
\begin{minipage}[b]{0.135\textwidth}
\includegraphics[width=1\textwidth]{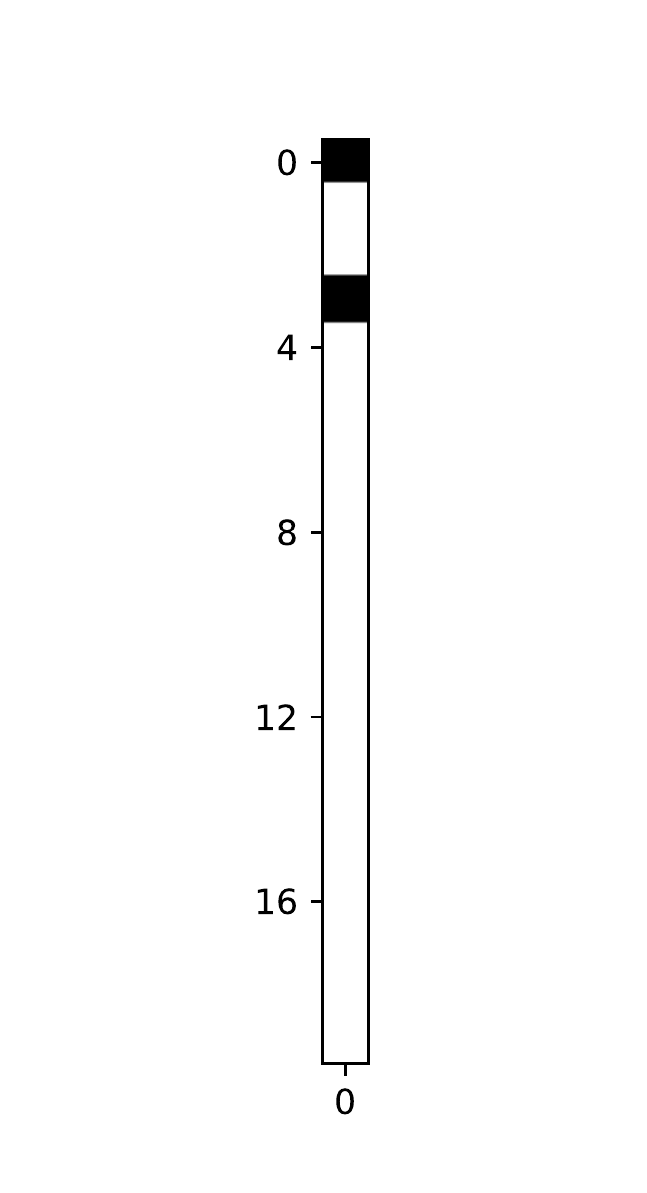}
\end{minipage}
}
\subfigure[]{
\begin{minipage}[b]{0.135\textwidth}
\includegraphics[width=1\textwidth]{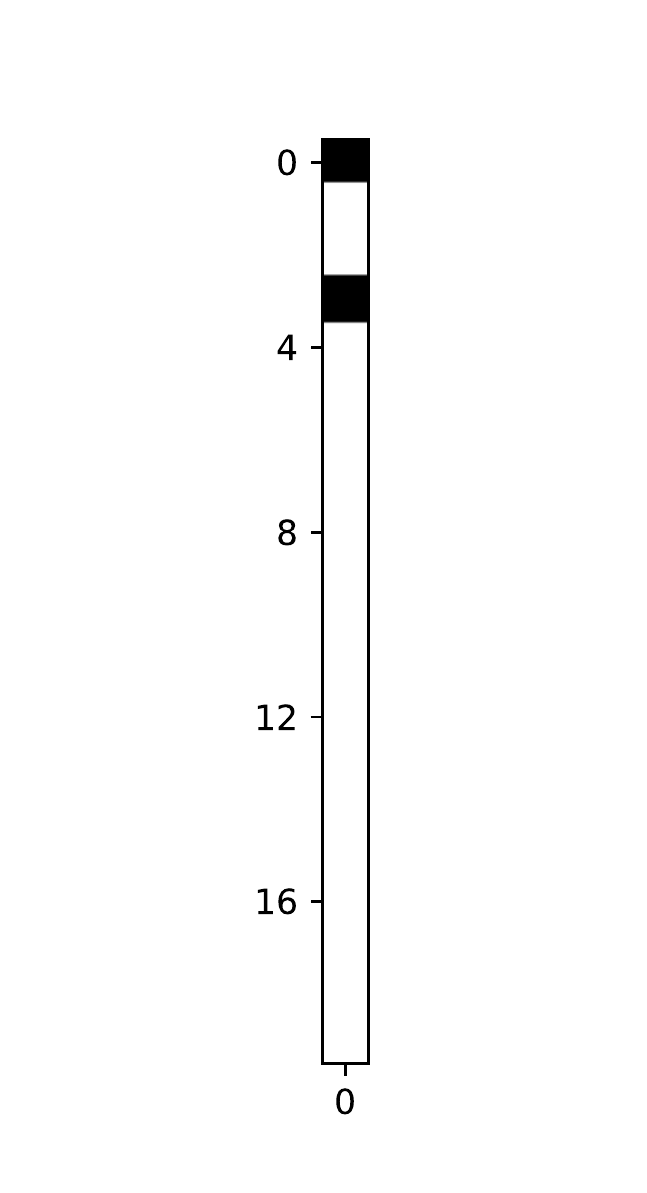}
\end{minipage}
}
\caption{A visualization of a well-trained DL-mAMPnet. (a) $\tilde{\bm{X}}_T$, the output of AMP layers; (b) The ground truth $\tilde{\bm{X}}$; (c) $\text{Mp}(\text{Mup}(\bm{o}))$, the output of the maxpool-
maxunpool procedure; (d) $\hat{\bm{\alpha}}$, the output of the refinement module; (e) The ground truth $\bm{\alpha}$.  }\label{vis}
\end{figure}

To offer more insights of the proposed DL-mAMPnet, we present the visualization of the outputs of each component of a well-trained DL-mAMPnet. For clarity, we only present the case where $N=10$ devices transmit 1-bit message with $\epsilon=0.1$ active probability, $L=10$ pilot sequence length, and $M=2$ receiving antennas. For visualization, we transform the outputs of each component to the reverse grayscale images. Specifically,
the elements of each output matrix are normalized to an interval within [0, 1], where 0 and 1 are represented by white color and black color, respectively. It should be mentioned that we take the absolute value of $\tilde{\bm{X}}$ and $\tilde{\bm{X}}_T$  to show the signal strength difference more intuitively.

The output of the AMP layers and its ground truth are shown in Fig.~\ref{vis}(a) and Fig.~\ref{vis}(b), respectively. It can be seen that the non-zero rows of $\tilde{\bm{X}}$ are correctly recovered, paving the way for the subsequent refinement progress. Then, the output of the maxpool-maxunpool procedure is visualized in Fig.~\ref{vis}(c), where the largest of the two adjacent rows is retained and the other becomes 0, demonstrating the validity of the maxpool-maxunpool procedure in ensuring the device-level sparsity. Fig.~\ref{vis}(d) and Fig.~\ref{vis}(e) are the visualizations of $\hat{\bm{\alpha}}$ and $\bm{\alpha}$, where we find that the pilot sequence activity is perfectly estimated by the well-trained DL-mAMPnet.  Moreover, it is observed from Fig.~\ref{vis}(c) and Fig.~\ref{vis}(e) that the pilot sequence activity is correctly reserved in Fig.~\ref{vis}(c) (the $1$st, $4$th, $21$st, and $24$th rows), which indicates the effectiveness of the proposed soft-thresholding denoising component.

\subsection{Computational Complexity Analysis}
Finally, we analyze the computational complexities of the  traditional AMP-based algorithm and DL-mAMPnet.

For the traditional AMP-based algorithm, the computational complexity mainly comes from the matrix multiplication in (\ref{amp1})-(\ref{amp2}) \cite{8}. Since $\bm{S}_n^H \in \mathbb{C}^{Q \times L}$, $\bm{R}_t \in \mathbb{C}^{L \times M}$,  $\bm{S} \in \mathbb{C}^{L \times NQ}$, and $\bm{X}_{t+1} \in \mathbb{C}^{NQ \times M}$, the computational complexity for $N$ devices and $T_{AMP}$ iterations is $\mathcal{O}(4T_{AMP}(NQLM+NQLM))=\mathcal{O}(8T_{AMP}NQLM)$, where the proportional constant ``$4$'' appears because a complex multiplication requires $4$ real multiplications, the former ``$NQLM$'' comes from the multiplication between $\bm{S}_n^{H}$ and $\bm{R}_t$ for $N$ devices and the latter ``$NQLM$'' comes from the multiplication between $\bm{S}$ and $\bm{X}_{t+1}$. After the iterative process, the AMP-based algorithm requires the element selection operation (i.e., (\ref{rm5})) whose computational complexity is $\mathcal{O}(4NQM)$, and the threshold calculation (i.e., (\ref{rm7}))  whose computational complexity is $\mathcal{O}(4NQM)$. Taking all the operations into account, the computational complexity of the AMP-based algorithm is given by $\mathcal{O}(8T_{AMP}NQLM)$.

For the proposed DL-mAMPnet, we focus on the computational complexity of online implementation. The  computational complexity of the AMP layers comes from the matrix multiplication $\bm{B}_{t}\tilde{\bm{R}}_{t-1}$ and $\bm{A}_t\tilde{\bm{X}}_{t}$, which is $\mathcal{O}(8T_{DL}NQLM^2)$ with $T_{DL}$ denoting the number of AMP layers.  For the refinement module, the  computational complexity is mainly resulted from the FC and convolutional layers. For a FC layer with $N_{l-1}$ input and $N_1$ output, its  computational complexity is given by $\mathcal{O}(N_{l-1}N_1)$. For a convolutional layer with a $H\times W$ input and a $H_f \times W_f$ filter, its  computational complexity can be expressed as $\mathcal{O}(HWH_fW_f)$. Therefore, the total computational complexity of the refinement module is $\mathcal{O}(4N^2Q^2M)$. Consequently, the  computational complexity of DL-mAMPnet is $\mathcal{O}(8T_{DL}NQLM^2+4N^2Q^2M)$.

From the above discussions, it seems that the proposed DL-mAMPnet can achieve better performance at the expense of a higher computational complexity compared to the AMP-based algorithm. However, as observed in Fig.~\ref{serl}-Fig.~\ref{serq}, the DL-mAMPnet with $T_{DL}=4$ AMP layers outperforms the AMP-based algorithm with $T_{AMP}=50$ iterations, indicating that the proposed DL-mAMPnet may need less computational complexity to achieve the same SER performance with the AMP-based algorithm.

\section{Conclusion}
This paper has proposed a novel DL-based algorithm, termed DL-mAMPnet, for the joint device activity and data detection in mMTC with a single-phase non-coherent scheme. Trainable parameters have been added in the DL-mAMPnet to compensate for the inaccuracy caused by the i.i.d. assumption in the traditional AMP algorithm. A refinement module has been further designed to enhance the SER performance and guarantee the device-level sparsity by exploiting the correlated sparsity pattern. The proposed algorithm can be applied to scenarios where massive users intermittently transmit small packets, e.g., smart home and industrial control. For the future work, we will investigate the pilot sequence design scheme to maintain orthogonality and mitigate the inter-device interference.

\begin{appendices}
\section{Derivation of MMSE Denoiser (\ref{amp8})}
To enable the derivation of the conditional probability $P_{\bm{X}_n|\bm{Z}_n}$, we assume $\bm{x}_n^q$ is independent with each other, and thus we have
\begin{equation}\label{ap1}
P_{\bm{x}_n^q} = \left(1-\frac{\epsilon}{Q} \right)\delta + \frac{\epsilon}{Q} \frac{\exp(-{\bm{x}_{n}^{q}}^{H}(\beta_n\bm{I}_M)^{-1}{\bm{x}_{n}^{q}})}{\pi^M|\beta_n\bm{I}_M|}.
\end{equation}

According to (\ref{amp7}), the likelihood of observing $\bm{z}_{n}^{q}$ given $\bm{x}_{n}^{q}$ is
\begin{equation}\label{ap2}
P_{\bm{z}_{n}^{q}|\bm{x}_{n}^{q}}=\frac{\exp (-(\bm{z}_{n}^{q}-\bm{x}_{n}^{q})^{H}\bm{\Sigma^{-1}}(\bm{z}_{n}^{q}-\bm{x}_{n}^{q}))}{\pi^M|\bm{\Sigma}|}.
\end{equation}

Denoting $k$ as the proportional constant, $P_{\bm{x}_{n}^{q}|\bm{z}_{n}^{q}}$ can be computed using the Bayes' formula as follows
\begin{align}\label{ap3}
&P_{\bm{x}_{n}^{q}|\bm{z}_{n}^{q}} = kP_{\bm{z}_{n}^{q}|\bm{x}_{n}^{q}} P_{\bm{x}_n^{q}} \nonumber \\
&= k  \left( (1-\frac{\epsilon}{Q})\delta + \frac{\epsilon}{Q} \frac{\exp(-{\bm{x}_{n}^{q}}^{H}(\beta_n\bm{I}_M)^{-1}{\bm{x}_{n}^{q}})}{\pi^M|\beta_n\bm{I}_M|} \right)  \left(\frac{\exp (-(\bm{z}_{n}^{q}-\bm{x}_{n}^{q})^{H}\bm{\Sigma^{-1}}(\bm{z}_{n}^{q}-\bm{x}_{n}^{q}))}{\pi^M|\bm{\Sigma}|}\right) \nonumber \\
&= k \left( (1-\frac{\epsilon}{Q})\frac{\exp (-{\bm{z}_{n}^{q}}^{H}\bm{\Sigma^{-1}} \bm{z}_{n}^{q})}{\pi^M|\bm{\Sigma}|}\delta  + \frac{\epsilon}{Q} \frac{\exp(-{\bm{x}_{n}^{q}}^{H}(\beta_n\bm{I}_M)^{-1}{\bm{x}_{n}^{q}}- (\bm{z}_{n}^{q}-\bm{x}_{n}^{q})^{H}\bm{\Sigma^{-1}}(\bm{z}_{n}^{q}-\bm{x}_{n}^{q}))}{\pi^{2M}|\beta_n\bm{I}_M||\bm{\Sigma}|} \right).
\end{align}

Note that
\begin{equation}\label{ap5}
{\bm{x}_{n}^{q}}^{H}(\beta_n\bm{I}_M)^{-1}{\bm{x}_{n}^{q}}+(\bm{z}_{n}^{q}-\bm{x}_{n}^{q})^{H}\bm{\Sigma^{-1}}(\bm{z}_{n}^{q}-\bm{x}_{n}^{q})\nonumber=(\bm{x}_{n}^{q}-\bm{\zeta})^{H}\bm{\Xi}^{-1}(\bm{x}_{n}^{q}-\bm{\zeta})+{\bm{z}_{n}^{q}}^{H}\bm{\Delta}^{-1}\bm{z}_{n}^{q},
\end{equation}
where $\bm{\Xi} = (\frac{1}{\beta_n}\bm{I}_M+ \bm{\Sigma}^{-1})$, $\bm{\zeta}=\bm{\Xi}\bm{\Sigma}^{-1}\bm{z}_{n}^{q}$, and $\bm{\Delta} = \beta_n\bm{I}_M +\bm{\Sigma}$, (\ref{ap3})  can be rewritten  as
\begin{align}\label{ap6}
&P_{\bm{x}_{n}^{q}|\bm{z}_{n}^{q}} \nonumber \\
&= k \left( (1-\frac{\epsilon}{Q})\frac{\exp (-{\bm{z}_{n}^{q}}^{H}\bm{\Sigma^{-1}} \bm{z}_{n}^{q})}{\pi^M|\bm{\Sigma}|}\delta + \frac{\epsilon}{Q} \frac{\exp \left(-(\bm{x}_{n}^{q}-\bm{\zeta})^{H}\bm{\Xi}^{-1}(\bm{x}_{n}^{q}-\bm{\zeta})-{\bm{z}_{n}^{q}}^{H}\bm{\Delta}^{-1}\bm{z}_{n}^{q} \right)}{\pi^{2M}|\beta_n\bm{I}_M||\bm{\Sigma}|} \right).
\end{align}

Since $\int P_{\bm{x}_{n}^{q}|\bm{z}_{n}^{q}}\, d\bm{x}_{n}^{q} =1 $, $k$ can be obtained by integrating (\ref{ap6})  out. Accordingly, we have
\begin{align}\label{ap7}
k &= \left( (1-\frac{\epsilon}{Q})\frac{\exp (-{\bm{z}_{n}^{q}}^{H}\bm{\Sigma^{-1}} \bm{z}_{n}^{q})}{\pi^M|\bm{\Sigma}|}  + \frac{\epsilon}{Q} \frac{\exp \left(-{\bm{z}_{n}^{q}}^{H}\bm{\Delta}^{-1}\bm{z}_{n}^{q}\right)|\bm{\Xi}|}{\pi^{M}|\beta_n\bm{I}_M||\bm{\Sigma}|} \right)^{-1} \nonumber \\
&\overset{(a)}{=}\left( (1-\frac{\epsilon}{Q})\frac{\exp (-{\bm{z}_{n}^{q}}^{H}\bm{\Sigma^{-1}} \bm{z}_{n}^{q})}{\pi^M|\bm{\Sigma}|}  + \frac{\epsilon}{Q} \frac{\exp \left(-{\bm{z}_{n}^{q}}^{H}\bm{\Delta}^{-1}\bm{z}_{n}^{q}\right)}{\pi^{M}|\bm{\Delta}|} \right)^{-1},
\end{align}
where $(a)$ holds because $|\frac{1}{\beta_n}\bm{I}_M+ \bm{\Sigma}^{-1}|=|\beta_n\bm{I}_M||\bm{\Sigma}|/|\beta_n\bm{I}_M+\bm{\Sigma}|$.

Substituting (\ref{ap7}) into (\ref{ap3}), $P_{\bm{x}_{n}^{q}|\bm{z}_{n}^{q}}$ can be determined as
\begin{equation}\label{ap8}
P_{\bm{x}_{n}^{q}|\bm{z}_{n}^{q}} = \frac{e^{-(\bm{x}_{n}^{q}-\bm{\zeta})^{H}\bm{\Xi}^{-1}(\bm{x}_{n}^{q}-\bm{\zeta})}\epsilon |\bm{\Sigma}|+ (Q-\epsilon) e^{-{\bm{z}_{n}^{q}}^{H}(\bm{\Sigma}^{-1}-\bm{\Delta}^{-1})\bm{z}_{n}^{q}} \pi^{M}|\bm{\Xi}| |\bm{\Delta}| \delta }{\epsilon \pi^{M}|\bm{\Xi}| |\bm{\Sigma}|+(Q-\epsilon)e^{-{\bm{z}_{n}^{q}}^{H}(\bm{\Sigma}^{-1}-\bm{\Delta}^{-1})\bm{z}_{n}^{q}} \pi^{M}|\bm{\Xi}||\bm{\Delta}|}.
\end{equation}

Hence, the conditional expectation $\mathbb{E}\{\bm{x}_{n}^{q}| \bm{z}_{n}^{q} \}$ is given by
\begin{align}\label{ap9}
\mathbb{E}\{\bm{x}_{n}^{q}| \bm{z}_{n}^{q} \} & = \int \bm{x}_{n}^{q} P_{\bm{x}_{n}^{q}|\bm{z}_{n}^{q}}\, d\bm{x}_{n}^{q}  = \frac{\bm{\zeta} \epsilon |\bm{\Sigma}|}{\epsilon |\bm{\Sigma}| +(Q-\epsilon)e^{-{\bm{z}_{n}^{q}}^{H}(\bm{\Sigma}^{-1}-\bm{\Delta}^{-1})\bm{z}_{n}^{q}} |\bm{\Delta}|}\nonumber \\
&= \frac{\beta_n(\beta_n \bm{I}_M + \bm{\Sigma})^{-1}\bm{z}_{n}^{q}}{1+\frac{Q-\epsilon}{\epsilon}|\bm{I}_M+\beta_n \bm{\Sigma}^{-1}|e^{-{\bm{z}_{n}^{q}}^{H}(\bm{\Sigma}^{-1}- (\bm{\Sigma}+\beta_n\bm{I}_M)^{-1})\bm{z}_{n}^{q}}}.
\end{align}

The MMSE-optimal denoiser in (\ref{amp8})-(\ref{amp12}) can be straightforwardly obtained from (\ref{ap9}) through simple mathematical transformation.

\end{appendices}

\linespread{1.2}

\linespread{1.2}

\begin{thebibliography}{99}
\bibitem{1}
N.~H.~Mahmood \emph{et al.}, ``White paper on critical and massive machine type communication towards 6G,'' \emph{arXiv preprint}, arXiv:2004.14146, 2020.
\bibitem{2}
X.~Chen, D.~W.~K.~Ng, W.~Yu, E.~G.~Larsson, N.~Al-Dhahir, and R.~Schober, ``Massive access for 5G and beyond,'' \emph{IEEE J. Sel. Areas Commun.}, vol.~39, no.~3, pp.~615--637, Mar.~2021.
\bibitem{3}
L.~Liu, E.~G.~Larsson, W.~Yu, P.~Popovski, C.~Stefanovic, and E.~de~Carvalho, ``Sparse signal processing for grant-free massive connectivity: A future paradigm for random access protocols in the Internet of things,'' \emph{IEEE Signal Process. Mag.}, vol.~35, no.~5, pp.~88--99, Sep.~2018.
\bibitem{4}
C.~Bockelmann, N.~Pratas, H.~Nikopour, K.~Au, T.~Svensson, C.~Stefanovic, P.~Popovski, and A.~Dekorsy, ``Massive machine-type communications in 5G: Physical and MAC-layer solutions,''  \emph{IEEE Commun. Mag.}, vol.~54, no.~9, pp.~59--65, Sep.~2016.
\bibitem{5}
M.~B.~Shahab, R.~Abbas, M.~Shirvanimoghaddam, and S.~J.~Johnson, ``Grant-free non-orthogonal multiple access for IoT: A survey,'' \emph{IEEE Commun. Surveys Tuts.}, vol.~22, no.~3, pp. 1805--1838, May~2020.
\bibitem{6}
E.~Bjornson, E.~de~Carvalho, J.~H.~Sorensen, E.~G.~Larsson, and P.~Popovski, ``A random access protocol for pilot allocation in crowded massive MIMO systems,'' \emph{IEEE Trans. Wireless Commun.}, vol.~16, no.~4, pp. 2220--2234, Apr.~2017.
\bibitem{7}
Z.~Chen, F.~Sohrabi, and W.~Yu, ``Sparse activity detection for massive connectivity,'' \emph{IEEE Trans. Signal Process.}, vol.~66, no.~7, pp.~1890--1904, Apr.~2018.
\bibitem{8}
L.~Liu and W.~Yu, ``Massive connectivity with massive MIMO-Part I: Device activity detection and channel estimation,'' \emph{IEEE Trans. Signal Process.}, vol.~66, no.~11, pp.~2933--2946, Jun.~2018.
\bibitem{9}
X.~Zhang, F.~Labeau, L.~Hao and J.~Liu, ``Joint active user detection and channel estimation via Bayesian learning approaches in MTC Communications,'' \emph{IEEE Trans. Veh. Technol.}, vol.~70, no.~6, pp.~6222--6226, Jun.~2021.
\bibitem{10}
J.~W.~Choi, B.~Shim, and S.-H.~Chang, ``Downlink pilot reduction for massive MIMO systems via compressed sensing,'' \emph{IEEE Commun. Lett.}, vol.~19, no.~11, pp.~1889--1892, Nov.~2015.
\bibitem{101}
T.~V.~Luong, Y.~Ko, N.~A.~Vien, M.~Matthaiou, and H.~Q.~Ngo, ``Deep energy autoencoder for noncoherent multicarrier MU-SIMO systems,'' \emph{IEEE Trans. Wireless Commun.}, vol.~19, no.~6, pp.~3952--3962, Jun.~2020.
\bibitem{102}
S.~Xue, Y.~Ma, and N.~Yi, "End-to-end learning for uplink MU-SIMO joint transmitter and non-coherent receiver design in fading channels," \emph{IEEE Trans. Wireless Commun.}, vol.~20, no.~9, pp.~5531--5542, Sep.~2021.
\bibitem{11}
K.~Senel and E.~G.~Larsson, ``Device activity and embedded information bit detection using AMP in massive MIMO,'' in \emph{Proc. IEEE GLOBECOM}, 2017, pp.~1--6.
\bibitem{12}
K.~Senel and E.~G.~Larsson, ``Joint user activity and non-coherent data detection in mMTC-enabled massive MIMO using machine learning algorithms,'' in \emph{Proc. WSA}, 2018, pp.~1--6.
\bibitem{13}
K.~Senel and E.~G.~Larsson, ``Grant-free massive MTC-enabled massive MIMO: A compressive sensing approach,'' \emph{IEEE Trans. Commun.}, vol.~66, no.~12, pp.~6164--6175, Dec.~2018.
\bibitem{14}
Z.~Chen, F.~Sohrabi, Y.-F.~Liu, and W.~Yu, ``Covariance based joint activity and data detection for massive random access with massive MIMO,'' in \emph{Proc. IEEE ICC}, 2019, pp.~1--6.
\bibitem{141}
X.~Shen, J.~Gao, W.~Wu, M.~Li, C.~Zhou, and W.~Zhuang, ``Holistic network virtualization and pervasive network intelligence for 6G,'' \emph{IEEE Commun. Surveys Tuts.}, vol.~24, no.~1, pp.~1--30, Firstquarter.~2022.
\bibitem{15}
X.~Shen \emph{et al.},``AI-assisted network-slicing based next-generation wireless networks,'' \emph{IEEE Open J. Veh. Technol.}, vol.~1, pp.~45--66, Jan.~2020.
\bibitem{16}
W.~Wu, N.~Cheng, N.~Zhang, P.~Yang, W.~Zhuang, and X.~Shen, ``Fast mmWave beam alignment via correlated bandit learning,'' \emph{IEEE Trans. Wireless Commun.}, vol.~18, no.~12, pp.~5894--5908, Dec.~2019.
\bibitem{17}
H.~He, C.~Wen, S.~Jin, and G.~Y.~Li, ``Model-driven deep learning for MIMO detection,'' \emph{IEEE. Trans. Signal Process.}, vol.~68, no.~1, pp.~1702--1715, Feb.~2020.
\bibitem{18}
M.~Soltani, V.~Pourahmadi, A.~Mirzaei, and H.~Sheikhzadeh, ``Deep learning-based channel estimation,'' \emph{IEEE Commun. Lett.}, vol.~23, no.~4, pp.~652--655, Apr.~2019.
\bibitem{19}
Z.~Ma, W.~Wu, M.~Jian, F.~Gao and X.~Shen, ``Joint constellation design and multiuser detection for grant-free NOMA,'' \emph{IEEE Trans. Wireless Commun.}, vol.~21, no.~3, pp.~1973--1988, Mar.~2022.
\bibitem{20}
J.~R.~Hershey, J.~Le~Roux, and F.~Weninger, ``Deep unfolding: Model-based inspiration of novel deep architectures,'' \emph{arXiv preprint}, arXiv:1409.2574, 2014.
\bibitem{21}
H.~He, S.~Jin, C.-K.~Wen, F.~Gao, G.~Y.~Li, and Z.~Xu, ``Model-driven deep learning for physical layer communications,'' \emph{IEEE Wireless Commun.}, vol.~26, no.~5, pp.~77--83, Oct.~2019.
\bibitem{22}
J.~Kim, W.~Chang, B.~Jung, D.~Baron, and J.~C.~Ye, ``Belief propagation for joint sparse recovery,''  \emph{arXiv preprint}, arXiv:1102.3289, 2011.
\bibitem{23}
J.~A.~Tropp, ``Algorithms for simultaneous sparse approximation. part II: Convex relaxation,'' \emph{Signal Processing}, vol.~86, no.~3, pp.~589--602, 2006.
\bibitem{24}
D.~L.~Donoho, A.~Maleki, and A.~Montanari, ``Message-passing algorithms for compressed sensing,'' \emph{Proc. Nat. Acad. Sci.} USA, vol.~106, no.~45, pp.~18914--18919, Nov.~2009.
\bibitem{25}
S. Rangan, ``Generalized approximate message passing for estimation with random linear mixing,''  in \emph{Proc. IEEE ISIT}, 2011, pp.~2168--2172.
\bibitem{26}
M.~Bayati and A.~Montanari, ``The dynamics of message passing on dense graphs, with applications to compressed sensing,'' \emph{IEEE Trans. Inf. Theory}, vol.~57, no.~2, pp.~764--785, Feb.~2011.
\bibitem{27}
M.~Borgerding, P.~Schniter, and S.~Rangan, ``AMP-inspired deep networks for sparse linear inverse problems,'' \emph{IEEE Trans. Signal Process.}, vol.~65, no.~16, pp.~4293--4308, Aug.~2017.
\bibitem{28}
D.~L.~Donoho, ``De-noising by soft-thresholding,'' \emph{IEEE Trans. Inf. Theory}, vol.~41, no.~3, pp.~613--627, May~1995.
\bibitem{29}
D.~Ciregan, U.~Meier, and J.~Schmidhuber, ``Multi-column deep neural networks for image classification,''  in \emph{Proc. IEEE CVPR}, 2012, pp.~3642--3649.
\bibitem{30}
X.~Gao \emph{et al.}, ``ComNet: Combination of deep learning and expert knowledge in OFDM receivers,'' \emph{IEEE Commun. Lett.}, vol.~22, no.~12, pp.~2627--2630, Dec.~2018.
\bibitem{31}
K.~He, X.~Zhang, S.~Ren, and J.~Sun, ``Delving deep into rectifiers: Surpassing human-level performance on ImageNet classification,'' in \emph{Proc. IEEE ICCV}, 2015, pp.~1026--1034.
\bibitem{32}
S.~K.~Kumar, ``On weight initialization in deep neural networks,'' \emph{arXiv preprint}, arXiv:1704.08863, 2017.
\bibitem{33}
C.~Metzler, A.~Mousavi, and R.~Baraniuk,``Learned D-AMP: Principled neural network based compressive image recovery,'' in \emph{Proc. NIPS}, 2017, pp.~1772--1783.
\end{thebibliography}
\end{document}